\documentclass[11pt]{article}
\usepackage{axodraw}
\usepackage{latexsym}
\usepackage{graphicx}
\usepackage{caption2}
\usepackage{psfrag}
\usepackage{amsmath}
\usepackage{supertabular}
\usepackage{longtable}
\input epsf

\begin{document}

\begin{titlepage}

\begin{flushright}
{
FTUV/07-05-21\\
IFIC/07-23\\
}
\end{flushright}
\vspace*{1.5cm}
\begin{center}
{\Large \bf Chiral Perturbation Theory with tensor sources}\\[2.0cm]

{\bf O. Cat\`{a}$^{a}$ and V. Mateu$^{b}$ }\\[1cm]

 $^{a}$Department of Physics, University of Washington\\ Seattle, WA 98195-1560, USA.\\[.5 cm]
 $^{b}$Departament de F\'isica Te\`orica, IFIC, Universitat de Val\`encia-CSIC\\ Apt. Correus 22085, E-46071 Val\`encia, Spain.

\end{center}

\vspace*{1.0cm}

\begin{abstract}
\noindent We construct the most general chirally invariant Lagrangian for mesons in the presence of external sources coupled to the tensor current ${\bar{\psi}}\,\sigma_{\mu\nu}\psi$. In order to have only even terms in the chiral expansion, we consider the new source of ${\cal{O}}(p^2)$. With this choice, we build the even-parity effective Lagrangian up to the $p^6$-order (NLO). While there are only 4 new terms at the $p^4$-order, at $p^6$-order we find 78 terms for $n_f=2$ and 113 terms for $n_f=3$. We provide a detailed discussion on the different mechanisms that ensure that our final set of operators is complete and non-redundant. We also examine the odd-parity sector, to conclude that the first operators appear at the $p^8$-order (NNLO).
\end{abstract}
\vfill
\hspace*{0.5cm} Keywords~: Chiral Lagrangians, Nonperturbative Effects,
Spontaneous Symmetry Breaking, QCD. \\ 
\eject
\end{titlepage}

\section{Introduction}

Chiral perturbation theory is the effective low-energy field theory of the strong interactions, describing the dynamics of pseudo-Goldstone bosons $(\pi, K, \eta)$ in an expansion in powers of external momenta and quark masses.
 
It was almost thirty years ago when Weinberg first introduced the chiral Lagrangian \cite{Wei}, in an effort to translate the current algebra relations of chiral symmetry into a Lagrangian formulation. This pioneering work was soon followed by the seminal papers of Gasser and Leutwyler \cite{Gass1,Gass2}, extending the chiral Lagrangian up to the next order in the chiral expansion (${\mathcal{O}}(p^4)$ in Weinberg's power counting scheme) and embedding it into a framework where Green's functions could be easily computed. In particular, they accounted for the set of Green's functions containing the Dirac bilinears ${\bar{\psi}}\,\gamma_{\mu}\psi$, ${\bar{\psi}}\,\gamma_{\mu}\gamma_5\psi$, ${\bar{\psi}}\,\psi$ and ${\bar{\psi}}\,i\gamma_5\psi$. 

Later on, and motivated by the increasing experimental accuracy, the Lagrangian was pushed to the two-loop order (${\cal{O}}(p^6)$) for the even-parity \cite{Fea,Bij} and odd-parity sectors \cite{Bijnens:2001bb,Ebertshauser:2001nj}. There is no theoretical difficulty in going to higher orders. Rather, the limitations come from the experimental inputs required to determine the growing number of free parameters.\footnote{Already at ${\cal{O}}(p^6)$, there are more than 100 low-energy couplings. Obviously, for certain physical processes only a few of these NNLO low-energy couplings need to be determined. It is for these processes that the two-loop computation is predictive.}

There have been efforts to extend the Lagrangian in the chiral expansion, but not in the number of external fields. In particular, no systematic introduction of the tensor Dirac bilinear ${\bar{\psi}}\sigma_{\mu\nu}\psi$ has been attempted. This can be partially explained because, as opposed to other Dirac currents, there is no physical realization of the tensor field coupled to the tensor Dirac bilinear in the Standard Model (SM). However, this does not mean that their phenomenology is uninteresting. For a long time several studies have turned their attention to Green's functions coupled to Dirac tensor sources, both from the sum rule perspective \cite{Iof}-\cite{Bak} and, more recently, from the lattice QCD perspective \cite{Bec}. Moreover, non-forward hadronic matrix elements of the Dirac tensor operator have recently attracted some attention because of their relevance in the study of generalized parton distributions (GPDs) of hadrons. In particular, the chiral realization of the tensor operator relevant for the GPDs of the pion was recently determined using the method of spurion fields (see, for instance, \cite{Diehl:2006js,Chen:2006gg} and references therein). A further phenomenological motivation for introducing such currents can be found when studying interactions beyond the SM. In certain scenarios one needs to calculate the hadronic matrix elements of tensor currents (see, for instance, \cite{vicentjorge}). It is the purpose of this paper to provide a consistent low-energy framework for such studies.

The paper is organized as follows~: in Section 2 we apply the external source method to the tensor sources coupled to the Dirac bilinear ${\bar{\psi}}\sigma_{\mu\nu}\psi$ and identify the list of basic  elements out of which we will construct the Lagrangian. Section 3 deals with the construction of the Lagrangian up to the $p^6$-order. We list the full set of operators invariant under Lorentz, chiral $SU(n_f)_L\times SU(n_f)_R$ and discrete symmetries, and reduce them to a minimal set through the use of the lowest-order equations of motion, integration by parts and Bianchi identities. If one specializes to the two-flavor and three-flavor cases, which are the phenomenologically relevant ones, further constraints are also provided by the Cayley-Hamilton relations. They are explicitly listed in the Appendix. Contact terms are discussed in Section 4. Section 5 is devoted to the odd-intrinsic parity sector, while in Section 6 we comment on the chiral counting for tensor sources. Finally, we give our conclusions in Section 7. Phenomenological applications are relegated to a companion paper.

\section{Preliminaries~: chiral building blocks}

Chiral perturbation theory ($\chi$PT) is the effective field theory describing the strong interactions at very low energies. It is based upon the global $SU(n_f)_L\times SU(n_f)_R$ flavor symmetry ($n_f=2,3$) spontaneously broken down to $SU(n_f)_V$.
The $\phi^a=(\pi,K,\eta)$ fields are assumed to be the Goldstone bosons of the theory and therefore their interactions are completely described by Goldstone dynamics. The general formalism for effective Lagrangians with spontaneously broken symmetries was worked out by Callan, Coleman, Wess and Zumino \cite{CWZ} and it is explained in detail for instance in Ref.~\cite{EFTManohar}. In particular, for our case it implies that the Goldstone bosons have to transform as
\begin{equation}
u(\phi^a)\rightarrow R\,u(\phi^a)\,h^{\dagger}\,=\,h\,u(\phi^a)\,L^{\dagger}\, ,
\end{equation}
where $R(L)\in SU(3)_{R(L)}$, $h \in SU(3)_V$ and $u(\phi^a)$ is a unitary non-linear representation of the Goldstone modes, typically
\begin{equation}
u(\phi^a)\,=\,{\mathrm{exp}}\left(\frac{i}{2F_0}\phi^a\lambda^a\right)\,.
\end{equation}
The chiral Lagrangian can then be built out of the invariant operators of $u(\phi^a)$ and its derivatives (or also commonly $U(\phi^a)\,=\,u(\phi^a)^2$ and its derivatives).

However, in order to compute Green's functions, it is convenient to introduce a set of external fields, one for each QCD current we want to account for, both in the QCD and $\chi$PT Lagrangians. Enforcement of chiral ward identities can then be easily achieved by promoting our global chiral symmetry to a local one \cite{Leut}. This is the basis of the external field method \cite{Gass1}, which we will briefly sketch below with the inclusion of the tensor external field. 

In terms of the effective action, the addition of external fields reads
\begin{eqnarray}
Z[v_{\mu},a_{\mu},s,p,{\bar{t}}_{\mu\nu}]&=&\int {\cal{D}}{\bar{\psi}}\,{\cal{D}}\psi\,{\cal{D}}G_{\mu} \,{\mathrm{exp}}\left[i\int d^4x\,\left\{ {\cal{L}}_{QCD}^0+{\cal{L}}_{ext}(v_{\mu},a_{\mu},s,p,{\bar{t}}_{\mu\nu})\right\}\right]\, ,\nonumber\\
&=& \int {\cal{D}}\,U{\cal{D}}U^{\dagger} \,{\mathrm{exp}}\left[i\int d^4x\,{\cal{L}}_{\chi}(U
;v_{\mu},a_{\mu},s,p,{\bar{t}}_{\mu\nu})\right]\, ,
\end{eqnarray}
where ${\cal{L}}^0_{QCD}$ is the massless QCD Lagrangian and
\begin{eqnarray}\label{sources}
{\cal{L}}_{ext}={\bar{\psi}}\,\gamma_{\mu}(v^{\mu}+\gamma_5\,a^{\mu})\,\psi-{\bar{\psi}}(s-i\,\gamma_5\,p)\,\psi+{\bar{\psi}}\,\sigma_{\mu\nu}\,{\bar{t}}^{\mu\nu}\,\psi\, ,
\end{eqnarray}
where $v_{\mu}$, $a_{\mu}$, $s$, $p$ and ${\bar{t}}^{\mu\nu}$ are hermitian matrices in flavor space. 
The vector and axial-vector external fields are chosen to be traceless in flavor space, but the rest of them will in general have a non-vanishing trace; for instance 
\begin{equation}
{\bar{t}}^{\mu\nu}\,=\,\sum_{a=0}^8\,\dfrac{\lambda^a}{2}\,{\bar{t}}^{\mu\nu}_a\,,
\end{equation} 
with $\lambda^0\,=\,\sqrt{2/n_f}\,{\bf{1}}_{n_f\times n_f}$. In order to manifestly show the chiral symmetry, it is convenient to first rotate our fields to the chiral basis, with projections given by
\begin{equation}\label{proj}
\psi_L\equiv P_L \,\psi=\left(\frac{1-\gamma_5}{2}\right)\psi\, , \qquad \psi_R\equiv P_R\, \psi=\left(\frac{1+\gamma_5}{2}\right)\psi \, ,
\end{equation}  
from which one can readily conclude that
\begin{equation}\label{ch-basis}
r_{\mu}\,=\,v_{\mu}+a_{\mu};\qquad l_{\mu}\,=\,v_{\mu}-a_{\mu};\qquad
\chi\,=\,2\,B_0\,(s+ip)\,,
\end{equation}
where $r_{\mu}$ and $l_{\mu}$ are coupled to right-handed and left-handed currents, respectively, while $\chi$ mixes the chiral sectors. $B_0$ is related to the quark condensate.

For the tensor field, one finds that
\begin{equation}
{\bar{\psi}}\,\sigma_{\mu\nu}\,{\bar{t}}^{\mu\nu}\psi\,=\,{\bar{\psi}}_L\sigma^{\mu\nu}t^{\dagger}_{\mu\nu}\psi_R+{\bar{\psi}}_R\,\sigma^{\mu\nu}t_{\mu\nu}\psi_L\, ,
\end{equation}
and the change of basis reads
\begin{eqnarray}
{\bar{t}}^{\mu\nu}&=&P_L^{\mu\nu\lambda\rho}\, t_{\lambda\rho}+P_R^{\mu\nu\lambda\rho}\, t^{\dagger}_{\lambda\rho}\, ,\nonumber\\
t^{\mu\nu}&=&P_L^{\mu\nu\lambda\rho}\,{\bar{t}}_{\lambda\rho}\,,\label{expr}
\end{eqnarray}
where $P_{L,R}^{\mu\nu\lambda\rho}$ are the analogs of $P_{L,R}$ in Eq.~(\ref{proj}) for the tensor fields, given by\footnote{In getting to (\ref{projectors}) use has been made of the algebraic identity
\begin{equation}
\sigma^{\mu\nu}\gamma_5\,=\,\frac{i}{2}\,\varepsilon^{\mu\nu\lambda\rho}\, \sigma_{\lambda\rho}\, .\nonumber
\end{equation}
The convention $\varepsilon^{0123}=+1$ for the Levi-Civit\'a tensor $\varepsilon^{\mu\nu\alpha\beta}$ will be used throughout this paper.}
\begin{eqnarray}\label{projectors} 
P_R^{\mu\nu\lambda\rho}&=&\frac{1}{4}(g^{\mu\lambda}g^{\nu\rho}-g^{\nu\lambda}g^{\mu\rho}+i\,\varepsilon^{\mu\nu\lambda\rho})\, ,\nonumber\\
P_L^{\mu\nu\lambda\rho}&=&\left( P_R^{\mu\nu\lambda\rho}\right)^\dagger \,.
\end{eqnarray}
Indeed, one can check that they satisfy the usual properties of chiral projectors
\begin{eqnarray}
P_R^{\mu\nu\lambda\rho}P_R^{\lambda\rho\alpha\beta}&=&P_R^{\mu\nu\alpha\beta}\, ,\nonumber\\ P_L^{\mu\nu\lambda\rho}P_L^{\lambda\rho\alpha\beta}&=&P_L^{\mu\nu\alpha\beta}\, ,\nonumber\\
P_L^{\mu\nu\lambda\rho}P_R^{\lambda\rho\alpha\beta}&=&0\, .
\end{eqnarray}
Eq.~(\ref{expr}) above just states the fact that $t_{\mu\nu}$ and $t^{\dagger}_{\mu\nu}$ are the left and right-handed projections of the tensor field and can be seen as the analog of Eq.~(\ref{ch-basis}). Notice that the chiral rotation mixes $v_{\mu}$ and $a_{\mu}$, $s$ and $p$ and the tensor with itself. This is precisely what one expects, since $\gamma_5$ acting on \textbf{$\sigma^{\mu\nu}$}
is not an independent Dirac matrix, but decomposable in terms of \textbf{$\sigma^{\mu\nu}$} alone. 
 
The next step is to promote the chiral symmetry to a local one. This sets the following chiral transformations for the various external fields
\begin{eqnarray}\label{sou}
r_{\mu}&\rightarrow& R\,r_{\mu}\,R^{\dagger}+iR\,\partial_{\mu}\,R^{\dagger}\, ,\nonumber\\
l_{\mu}&\rightarrow& L\,l_{\mu}\,L^{\dagger}+iL\,\partial_{\mu}\,L^{\dagger}\, ,\nonumber\\
\chi &\rightarrow& R\,\chi\, L^{\dagger}\, ,\nonumber\\
t_{\mu\nu} &\rightarrow& R\, t_{\mu\nu}\, L^{\dagger}\, ,
\end{eqnarray}
together with a covariant derivative for the pion field, namely
\begin{eqnarray}\label{cov}
D_{\mu}U=\partial_{\mu}U-i\,r_{\mu}\,U+i\,U\,l_{\mu}\, &,& D_{\mu}U\rightarrow R\,(D_{\mu}U)\,L^{\dagger}\, ;\nonumber\\
D_{\mu}U^{\dagger}=\partial_{\mu}U^{\dagger}+i\,U^{\dagger}\,r_{\mu}-i\,l_{\mu}\,U^{\dagger}\, &,& D_{\mu}U^{\dagger}\rightarrow L\,(D_{\mu}U^{\dagger})\,R^{\dagger}\, .
\end{eqnarray}
For the right and left-handed fields, field strength tensors arise naturally~:
\begin{equation}\label{strength}
[D^{\mu},D^{\nu}]\,X\,=\,i\,XF_L^{\mu\nu}-i\,F_R^{\mu\nu}X\, ,
\end{equation}
with 
\begin{equation}\label{Faraday}
F_L^{\mu\nu}\,=\,\partial^{\mu}l^{\nu}-\partial^{\nu}l^{\mu}-i\,[l^{\mu},l^{\nu}]\, , \quad F_R^{\mu\nu}\,=\,\partial^{\mu}r^{\nu}-\partial^{\nu}r^{\mu}-i\,[r^{\mu},r^{\nu}]\, .
\end{equation}
The set $(U,F_{L,R}^{\mu\nu},\chi,t_{\mu\nu})$, along with their adjoints and covariant derivatives, are the building blocks to construct a theory with chiral symmetry. The next step would be to assemble them together in chiral invariant combinations which respect parity, charge conjugation and hermiticity. 

However, the building blocks listed above transform differently under the chiral group. This is not a problem when one is dealing with the lowest orders in the chiral expansion, where the combinatorics are simple and only a small number of operators result. However, already at next-to-leading order the number of operators involved recommends to deal with building blocks in a more efficient way. We will follow the conventions of \cite{Bij,Eckerres} and work with the following set of hermitian and anti-hermitian terms,
\begin{eqnarray}\label{listinv}
u_{\mu}&=&i\left\{ u^{\dagger}(\partial_{\mu}-ir_{\mu})u-u(\partial_{\mu}-il_{\mu})u^{\dagger}\right\} \equiv i\,u^{\dagger}\,D_{\mu}U\,u^{\dagger}\, ,\nonumber\\
h_{\mu\nu}&=&\nabla_{\mu}u_{\nu}+\nabla_{\nu}u_{\mu}\, ,\nonumber\\
f_{\pm}^{\mu\nu}&=&u\,F_L^{\mu\nu}\,u^{\dagger}\, \pm \, u^{\dagger}\, F_R^{\mu\nu}\, u\, ,\nonumber\\
t_{\pm}^{\mu\nu}&=&u^{\dagger}\, t^{\mu\nu}\, u^{\dagger}\, \pm \, u\, t^{\mu\nu\,\dagger} \, u\, ,\nonumber\\
\chi_{\pm}&=&u^{\dagger} \, \chi \, u^{\dagger}\, \pm \, u\, \chi^{\dagger}\, u\, ,
\end{eqnarray}
where signs are correlated. The main advantage of this new set of operators is that they all transform in the same manner under the chiral group, namely
\begin{equation}
h\, X\, h^{\dagger}\, , \qquad X=u_{\mu}, f_{\pm}^{\mu\nu}, t_{\pm}^{\mu\nu},...
\end{equation} 
As a result, one can define a unique covariant derivative for them all, {\it{e.g.}},
\begin{equation}\label{conn}
\nabla_{\rho}X=\partial_{\rho}X+[\Gamma_{\rho},X]\, , \qquad \Gamma_{\rho}=\frac{1}{2}\left\{u^{\dagger}(\partial_{\rho}-ir_{\rho})u+u(\partial_{\rho}-il_{\rho})u^{\dagger}\right\}\, ,
\end{equation} 
where the last term is the chiral connection. Analogously to Eq.~(\ref{cov}), there is a field strength tensor associated to the covariant derivative, namely
\begin{equation}\label{fstrength}
\left[\nabla_{\mu},\nabla_{\nu}\right]\, X =  \left[\Gamma_{\mu\nu},\, X\right],
\end{equation}
with
\begin{equation}\label{faraday}
\Gamma_{\mu\nu} = \partial_{\mu}\Gamma_{\nu}-\partial_{\nu}\Gamma_{\mu}+\left[\Gamma_{\mu},\Gamma_{\nu}\right]=\frac{1}{4}\left[u_{\mu},u_{\nu}\right]-\frac{i}{2}f_{+\mu\nu}\, .
\end{equation}
Both the sets $(U^{(\dagger)},F_{L,R}^{\mu\nu},\chi^{(\dagger)},t^{(\dagger)}_{\mu\nu})$ and $(u_{\mu},h_{\mu\nu},f_{\pm}^{\mu\nu},\chi_{\pm},t^{\mu\nu}_{\pm})$ are complete.\footnote{Note that $u_{\mu}$ is self-adjoint and the combination $\nabla^{\nu}u^{\mu}-\nabla^{\mu}u^{\nu}=f_{-}^{\mu\nu}$ and therefore it is redundant.} Therefore, both can be used to build the chiral Lagrangian. The latter set, as mentioned previously, eases the path to determining the full set of operators in the chiral Lagrangian and will be adopted in the next section. The former set, however, has no mixing between Goldstone modes and external fields and it will prove useful in Section 4, when we will isolate contact terms.\renewcommand{\arraystretch}{1.3}
\begin{table}
\begin{center}
\begin{tabular}{|c|c|c|c|}
\hline
$\mathcal{O}$ & $\mathcal{P}$ & $\mathcal{C}$ & h.c. \\
\hline
$u^{\mu}$ & $-u_{\mu}$  & $(u^{\mu})^T$ & $u^{\mu}$  \\
$h^{\mu\nu}$ & $-h_{\mu\nu}$  & $(h^{\mu\nu})^T$ & $h^{\mu\nu}$  \\
$\chi_{\pm}$ & $\pm\chi_{\pm}$  & $(\chi_{\pm})^T$ & $\pm \chi_{\pm}$  \\
$f_{\pm}^{\mu\nu}$ & $\pm f_{\pm\mu\nu}$  & $\mp (f_{\pm}^{\mu\nu})^T$ & $ f_{\pm}^{\mu\nu}$  \\
$t_{\pm}^{\mu\nu}$ & $\pm t_{\pm\mu\nu}$ &$-(t_{\pm}^{\mu\nu})^T$  & $\pm t_{\pm}^{\mu\nu}$  \\
\hline
\end{tabular}
\end{center} \caption{Various transformation properties of the elements of Eq.~(\ref{listinv}). \label{discrete}}\end{table}

\section{Construction of the Effective Lagrangian}
The whole set of operators including tensor sources can now be built by assembling together the building blocks of Eq.~(\ref{listinv}) (and their covariant derivatives) in traces and products of traces thereof with the help of Table \ref{discrete}, such that the Lagrangian be hermitian and invariant under discrete symmetries.
 
Prior to the actual construction of the Lagrangian, however, all (external) fields have to be endowed with a power counting, such that the resulting operators can be accommodated in the chiral expansion. In order to have only even terms in the chiral expansion, it is convenient to choose 
\begin{eqnarray}
U=u^2 &\sim& {\cal{O}}(p^0)\, ,\nonumber\\
v_{\mu},a_{\mu}&\sim& {\cal{O}}(p^1)\, ,\nonumber\\
\chi &\sim& {\cal{O}}(p^2)\, ,\nonumber\\
t_{\mu\nu} &\sim& {\cal{O}}(p^2)\, .
\end{eqnarray}
With these conventions, operators with tensor fields appear first at ${\cal{O}}(p^4)$. 
\subsection{Chiral Lagrangian to lowest order}
When the external tensor field is switched on, one finds the following operators
\begin{equation}\label{O4}
{\cal{L}}_4^{\chi PT}=\Lambda_1\,\langle\, t_+^{\mu\nu}\,f_{+\mu\nu}\,\rangle\,-\,i\,\Lambda_{2}\,\langle\, t_+^{\mu\nu}\,u_{\mu}u_{\nu}\,\rangle\,+\,\Lambda_{3}\,\langle\, t_+^{\mu\nu}\,t_{\mu\nu}^+\,\rangle\,+\,\Lambda_{4}\,\langle\, t_+^{\mu\nu}\,\rangle^2\, ,
\end{equation}
where $\langle\, \cdots \rangle$ stands for the trace in $n_f$ flavor space. In Eq.~(\ref{O4}) and all through our analysis we will make extensive use of the tracelessness properties $\langle r_{\mu}\rangle=0=\langle F_R^{\mu\nu}\rangle$, $\langle l_{\mu}\rangle=0=\langle F_L^{\mu\nu}\rangle$ and $\langle u_{\mu}\rangle=0=\langle f_{\pm}^{\mu\nu}\rangle$.

It is interesting to remark that a potential contact term like $t^{\mu\nu}\,t^{\dagger}_{\mu\nu}$ in Eq.~(\ref{O4}) cancels identically due to orthogonality of chiralities, as can be easily checked using the chiral projectors of Eq.~(\ref{expr}).\footnote{Obviously, the tensor source is not a Lorentz scalar, and contact terms involving det $(t_{\mu\nu})$ are forbidden.} Hence, it follows that $t_+^{\mu\nu}\,t^+_{\mu\nu}=t_-^{\mu\nu}\,t^-_{\mu\nu}$ and $t_+^{\mu\nu}\,t^-_{\mu\nu}=t_-^{\mu\nu}\,t^+_{\mu\nu}$. These relations have been used in deriving Eq.~(\ref{O4}) and will be used hereafter.

We also note that Eq.~(\ref{O4}) is valid for any number of flavors. Cayley-Hamilton trace relations (to be discussed in the Appendix) do not provide extra constraints.

\subsection{Higher order terms}

At next-to-leading order (${\cal{O}}(p^6)$), the number of operators with tensor sources increases considerably. The purpose of this section is to sketch the steps followed in reaching the basis of chiral invariant operators listed in Table \ref{list2}. In particular, we will outline the strategies followed to reduce the set of operators to a non-redundant minimal one, focussing on the results obtained rather than giving the technical details, which can be found in \cite{Fea,Bij}.

The full set of ${\cal{O}}(p^6)$ operators which results from combining the building blocks of Eq.~(\ref{listinv}) and their covariant derivatives falls into one of the following generic groups
\begin{eqnarray}\label{prelim}
t_{\mu\nu}\,t_{}^{\mu\nu}\,u_{\alpha}\,u^{\alpha}\,;\quad
&t_{\mu\nu}\,f_{}^{\mu\nu}\,\chi_{}\,;&
\quad t_{\mu\nu}\,t_{}^{\mu\nu}\,\chi_{}\,;\nonumber\\
t_{\mu\nu}\,\chi_{}\,u^{\mu}\,u^{\nu}\,;\quad
&t_{\mu\nu}\,f_{}^{\mu\rho}\,f_{\,\,\rho}^{\nu}\,;&
\quad t_{\mu\nu}\,t_{}^{\nu\rho}\,h_{\,\,\rho}^{\mu}\,;\nonumber\\
t_{\mu\nu}\,h^{\nu\rho}\,u_{\rho}\,u^{\mu}\,;\quad
&t_{\mu\nu}\,h^{\mu\alpha}\,h_{\,\,\alpha}^{\nu}\,;&
\quad t_{\mu\nu}\,t_{}^{\nu\rho}\,f_{\,\,\rho}^{\mu}\,;\nonumber\\
t_{\mu\nu}\,f_{}^{\mu\nu}\,u_{\alpha}\,u^{\alpha}\,;\quad
&t_{}^{\mu\nu}\,\chi_{\mu}\,u_{\nu}\,;&
\quad t_{\mu\nu}\,t_{}^{\mu\rho}\,t_{\,\,\rho}^{\nu}\,;\nonumber\\
\nabla_{\rho}\,t_{\mu\nu}\,\nabla^{\rho}\,t^{\mu\nu}\,;\quad
&t_{\mu\nu}\,h^{\mu\alpha}\,f_{\,\,\alpha}^{\nu}\,;&
\quad \nabla_{\mu}\,t_{}^{\mu\nu}\,\nabla^{\alpha}\,f_{\alpha\nu}\,;\nonumber\\
\nabla_{\rho}\,t_{\mu\nu}\,h^{\mu\rho}\,u^{\nu}\,;\quad
&\nabla^{\mu}\,t_{\mu\nu}\,f_{}^{\nu\rho}\,u_{\rho}\,;&
\quad \nabla_{\lambda}\,t_{}^{\mu\nu}\,t_{\mu\nu}\,u^{\lambda}\,;\nonumber\\
&t_{\mu\nu}\,u_{\alpha}\,u^{\mu}\,u^{\nu}\,u_{\alpha}\,;&
\end{eqnarray}
where emphasis has been placed only on operator combinations, {\it{i.e.}}, traces and $i$ factors have been omitted and $\pm$ subscripts have been skipped for simplicity. Also, we have used the short-hand notation $\nabla_{\mu}\,\chi\equiv \chi_{\mu}$. The previous list is however complete in the sense that it contains all the independent operator combinations. For instance, operators like $\nabla^{\lambda}\,t_{\mu\nu}\,\chi_{\lambda}$ are generically C-violating and $\nabla^{\mu}\,t_{\mu\nu}\,u^{\nu}\,u_{\alpha}\,u^{\alpha}$ or $t^{\mu\nu}\,\nabla^{\lambda}f_{\mu\nu}\,u_{\lambda}$ can be shown to be redundant using partial integration and the chain rule. 

Table \ref{list2} lists the full set of hermitian operators invariant under parity and charge conjugation, organized in blocks of operators below each of the representatives of Eq.~(\ref{prelim}).

Obviously, the most challenging task in going from Eq.~(\ref{prelim}) above to our final set of operators in Table \ref{list2} is to make sure that the set is minimal, {\it{i.e.}}, linearly dependent operators have been removed and we can talk of a true chiral basis of operators. In the following we will discuss the commonly used strategies, namely integration by parts, use of the equations of motion\footnote{In determining the higher order terms in the chiral expansion the equations of motion for the leading order can be used. As discussed in \cite{Bij}, its enforcement is equivalent to a transformation of fields and therefore physics is left invariant.} and the Bianchi identity.
\subsubsection{Partial integration and equations of motion}
Integration by parts was already used to get to Eq.~(\ref{prelim}). The list can be further reduced, however, if one notices that the covariant derivatives of $u_{\mu}$ satisfy
\begin{equation}\label{rel}
\nabla_{\mu}u_{\nu}=\frac{1}{2}\left(\,h_{\mu\nu}-f_{-\mu\nu}\,\right)\, .
\end{equation}
Furthermore, the lowest-order equations of motion for the chiral Lagrangian read
\begin{equation}\label{EOM}
\nabla_{\mu}u^{\mu}=\frac{1}{2\,i}\left(\frac{\langle \chi_-\rangle}{n_f}-\chi_-\right)\, .
\end{equation}
If we combine Eqs.~(\ref{rel}) and (\ref{EOM}) with integration by parts we find the following relations,
\begin{eqnarray}
 i\left\{\nabla_{\lambda}\,t_-^{\mu\nu}\,,\,t_{+\mu\nu}\,\right\}\, u^{\lambda}&=& -i\left\{ \nabla_{\lambda}\,t_+^{\mu\nu}\,,\,t_{-\mu\nu} \right\}\, u^{\lambda}\,+\, t_+^{\mu\nu}\,t_{-\mu\nu}\, \chi_-\,-\,\frac{1}{n_f}\, t_+^{\mu\nu}\,t_{-\mu\nu}\, \langle \,\chi_-\,\rangle\, , \nonumber\\
i\left\{ \nabla_{\mu}\,t_-^{\mu\nu}\,,\,t_{+\nu\lambda}\, \right\}\, u^{\lambda}&=&-i\left\{ \nabla^{\lambda}\,t_{+}^{\mu\nu}\,,\,t_{-\nu\lambda}\, \right\}\, u_{\mu}\,+\,\frac{i}{2}\left\{\,t_{+\nu}^{\mu}\,,\,t_{-\mu\alpha}\,\right\}\, h^{\nu\alpha}\,+\,\frac{i}{2} \left\{\, t_{+\nu}^{\mu}\,,\,t_{-\mu\alpha}\, \right\}\, f_{-}^{\nu\alpha}\, ,\nonumber\\
i\left\{ \nabla_{\mu}\,t_+^{\mu\nu}\,,\,t_{-\nu\lambda}\, \right\}\, u^{\lambda}&=& -i\left\{ \nabla^{\lambda}\,t_-^{\mu\nu}\,,\,t_{+\nu\lambda}\, \right\}\, u_{\mu}\,+\,\frac{i}{2}\left\{\,t_{+\nu}^{\mu}\,,\,t_{-\mu\alpha}\,\right\}\, h^{\nu\alpha}\,-\,\frac{i}{2}\left\{\, t_{+\nu}^{\mu}\,,\,t_{-\mu\alpha}\, \right\}\, f_{-}^{\nu\alpha}\, ,\nonumber\\
&&
\end{eqnarray}
where in the first line the lowest-order equations of motion of Eq.~(\ref{EOM}) have been used. The second and third relations follow from Eq.~(\ref{rel}).
  
Further relations can be found using the properties of the chiral connection listed in Eqs.~(\ref{fstrength}) and (\ref{faraday}). In particular,
\begin{eqnarray}
\nabla^{\lambda}\,t^+_{\lambda\nu}\, \nabla_{\rho}\,t^{\rho\nu}_+\,-\,\nabla^{\rho}\,t^+_{\lambda\nu}\, \nabla^{\lambda}\,t^{\rho\nu}_+&=&[\,\Gamma^{\lambda\rho}\,,\,t^{+\nu}_{\lambda}\,]\,t^+_{\rho\nu}\,\,=\,\,-\frac{1}{2}\,Y_{11}\,+\,\frac{1}{2}\,Y_{12}\,+\,Y_{90}\, ,\nonumber\\
\nabla^{\lambda}\,t^-_{\lambda\nu}\, \nabla_{\rho}\,t^{\rho\nu}_- \,-\, \nabla^{\rho}\,t^-_{\lambda\nu}\, \nabla^{\lambda}\,t^{\rho\nu}_-&=&[\,\Gamma^{\lambda\rho}\,,\,t^{-\nu}_{\lambda}\,]\,t^-_{\rho\nu}\,\,=\,\,-\frac{1}{2}\,Y_{23}\,+\,\frac{1}{2}\,Y_{24}\,+\,Y_{91}\, ,\nonumber\\
\nabla^{\lambda}\,t^+_{\lambda\nu}\, \nabla_{\rho}\,f^{\rho\nu}_+ \,-\, \nabla^{\rho}\,t^+_{\lambda\nu}\, \nabla^{\lambda}\,f^{\rho\nu}_+&=&[\,\Gamma^{\lambda\rho}\,,\,t^{+\nu}_{\lambda}\,]\,f^+_{\rho\nu}\,\,=\,\,\frac{1}{4}\,Y_{59}\,-\,\frac{1}{4}\,Y_{60}\,-\,Y_{85}\, ,
\end{eqnarray}
where the $Y_i$ operators can be found in Table \ref{list2}. We have chosen to eliminate the second operators in the left-hand side in the equations above. In a similar fashion (but after a more involved calculation), one can show that $i\,\left\langle\, \nabla_{\rho}\,t_{+\mu\nu}\,\left\{\, f_{-}^{\mu\nu}\,,\,u^{\rho}\,\right\}\, \right\rangle $ and $i\,\left\langle\, \nabla_{\rho}\,t_{+\mu\nu}\,\left\{\, f_{-}^{\mu\rho}\,,\,u^{\nu}\,\right\} \,\right\rangle$ are also redundant.

\subsubsection{Bianchi identity}
In Eqs.~(\ref{conn})-(\ref{faraday}) we introduced the chiral connection and the field strength $\Gamma_{\mu\nu}$ that naturally stems from it. There is also an associated Bianchi identity, which in this case takes the form
\begin{equation}\label{Bianchi}
\nabla_{\mu}\,\Gamma_{\nu\rho}\,+\,\nabla_{\nu}\,\Gamma_{\rho\mu}\,+\,\nabla_{\rho}\,\Gamma_{\mu\nu}\,=\,0.
\end{equation}
Tracing this equation with $\nabla^{\rho}\,t_{+}^{\mu\nu}$ and integrating
by parts we get one additional relation between operators. Accordingly, we choose to remove from our list the operator $i\left\langle \nabla^{\rho}\,t_{+\mu\nu}\nabla_{\rho}\,f_{+}^{\mu\nu}\right\rangle$.
\section{Contact terms}
So far, to the best of our knowledge the number of operators for general $n_f$ is complete and minimal. However, in our list there are contact terms, {\it{i.e.}}, combinations of operators which only depend on external sources. Since they do not contain the pion field, they cannot be determined from phenomenology, but are necessary to correctly account for the ultraviolet behavior of Green's functions.

In the basis or hermitian and anti-hermitian chiral invariants we have been using, contact terms do not arise in a natural way. Instead, they are hidden in linear combinations of operators. As we already discussed, chirality prevents a contact term like $t^{\mu\nu}t_{\mu\nu}^{\dagger}$ at order ${\cal{O}}(p^4)$. At the next order, one finds the following contact terms
\begin{eqnarray}
\left\langle D_{\mu}t^{\mu\nu}D^{\alpha}t_{\alpha\nu}^{\dagger}\right\rangle  & = & \frac{1}{4}\left\langle \nabla_{\mu}t_{+}^{\mu\nu}\nabla^{\alpha}t_{+\alpha\nu}\right\rangle -\frac{1}{4}\left\langle \nabla_{\mu}t_{-}^{\mu\nu}\nabla^{\alpha}t_{-\alpha\nu}\right\rangle -\nonumber\\
 & - & \frac{i}{4}\left\langle \nabla_{\mu}t_{+}^{\mu\nu}\left\{ t_{-\alpha\nu},u^{\alpha}\right\} \right\rangle +\frac{1}{16}\left\langle t_{+\mu\nu}\left(u^{\nu}t_{+}^{\mu\alpha}u_{\alpha}+u_{\alpha}t_{+}^{\mu\alpha}u^{\nu}\right)\right\rangle +\nonumber\\
 & + & \frac{1}{8}\left\langle t_{+\mu\nu}t_{+}^{\mu\alpha}u_{\alpha}u^{\nu}\right\rangle -\frac{1}{16}\left\langle t_{-\mu\nu}\left(u^{\nu}t_{-}^{\mu\alpha}u_{\alpha}+u_{\alpha}t_{-}^{\mu\alpha}u^{\nu}\right)\right\rangle -\nonumber\\
 & - & \frac{1}{8}\left\langle t_{-\mu\nu}t_{-}^{\mu\alpha}u_{\alpha}u^{\nu}\right\rangle +\frac{i}{4}\left\langle \nabla_{\mu}t_{-}^{\mu\nu}\left\{ t_{+\alpha\nu},u^{\alpha}\right\} \right\rangle \,,\nonumber\\ 
\left\langle 
t^{\dagger \nu\rho}t^{\mu}_{\,\,\rho} F_{L \mu\nu}+
t^{\nu\rho}t^{\dagger\mu}_{\,\,\,\,\,\rho} F_{R \mu\nu} 
\right\rangle 
& = &
\frac{1}{4}\left\langle t_{+}^{\nu\rho}t_{+\rho}^{\mu}f_{+\mu\nu}\right\rangle-\frac{1}{4}\left\langle t_{-}^{\nu\rho}t_{-\rho}^{\mu}f_{+\mu\nu}\right\rangle +\frac{1}{4}\left\langle \left\{ t_{+}^{\nu\rho},t_{-\rho}^{\mu}\right\} f_{-\mu\nu}\right\rangle\, ,\nonumber\\
\left\langle t_{\mu\nu}\chi^{\dagger}F_{R}^{\mu\nu}+
\chi^{\dagger}t_{\mu\nu}F_{L}^{\mu\nu}+{\mathrm{h.c.}}\right\rangle  & = & \frac{1}{4}\left\langle t_{+\mu\nu}\left\{ f_{+}^{\mu\nu},\chi_{+}\right\} \right\rangle +\frac{1}{4}\left\langle t_{-\mu\nu}\left[f_{-}^{\mu\nu},\chi_{+}\right]\right\rangle -\nonumber\\
 & - & \frac{1}{4}\left\langle t_{-\mu\nu}\left\{ f_{+}^{\mu\nu},\chi_{-}\right\} \right\rangle -\frac{1}{4}\left\langle t_{+\mu\nu}\left[f_{-}^{\mu\nu},\chi_{-}\right]\right\rangle\, ,\end{eqnarray}
where $\nabla_{\mu}t^{\mu\nu}_{\pm}=(D_{\mu}t^{\mu\nu})_{\pm}+\frac{i}{2}\left\{u_{\mu},t^{\mu\nu}_{\mp}\right\}$ has been used in the first relation. We will incorporate the previous contact terms in our basis, and accordingly remove the following monomials, which otherwise would be redundant~:
\begin{eqnarray}
i \left\langle \left\{ t_{+}^{\nu\rho},t_{-\rho}^{\mu}\right\} f_{-\mu\nu}\right\rangle  & = & -\,Y_{90}+Y_{91}+4\,Y_{119}\,,\nonumber\\
i\left\langle \nabla_{\mu}t_{-}^{\mu\nu}\left\{ t_{+\alpha\nu},u^{\alpha}\right\} \right\rangle & = & -\,\frac{1}{2}\,Y_{11}-\frac{1}{4}\,Y_{13}+\frac{1}{2}\,Y_{23}+\frac{1}{4}\,Y_{25}-Y_{52}+Y_{53}-Y_{105}+4\,Y_{118}\,,\nonumber\\
\left\langle t_{-\mu\nu}\left[f_{-}^{\mu\nu},\chi_{+}\right]\right\rangle  & = & -\,Y_{74}+Y_{75}+Y_{76}+4\,Y_{120}\,.\end{eqnarray}
All the relations discussed above finally reduce the number of operators to 117 and 3 contact terms. This is the number of independent operators for any number of flavors. However, only $n_f=2,3$ are phenomenologically relevant. For such cases, the Cayley-Hamilton theorem provides further relations between traces. For reference, we list them in the Appendix. After enforcing the Cayley-Hamilton relations, we end up with 110+3 independent operators for three flavors and 75+3 for two flavors.

In order to have a minimal basis of $\mathcal{O}(p^6)$ chirally invariant monomials with tensor sources, we have followed the same procedure as in Ref.~\cite{Bij}. However, a recent paper \cite{Haefeli:2007ty} has pointed out that the basis of \cite{Bij} for two flavors is not yet minimal~: an identity among several operators of that basis was found, which does not become trivial when setting to zero the external sources. Interestingly, such identity does not require new algebraic manipulations other than the Cayley-Hamilton relations, Bianchi identities, partial integration and equations of motion. The fact that even after the sophisticated analysis of \cite{Bij} an additional relation was found shows that reaching a minimal set of operators at higher orders in the chiral expansion is quite a challenging task. With tensor sources, however, highly nontrivial relations such as the one reported in Ref.~\cite{Haefeli:2007ty} are unlikely to be found, mainly because~: (a) the tensor source does not enter the lowest order equations of motion and (b) there is no Bianchi identity associated with it. As a result, algebraic manipulations are simpler and we do not expect our basis to suffer further reduction. 

\section{A comment on the odd-intrinsic-parity sector}
So far we have restricted our analysis to the even-intrinsic parity sector of the chiral expansion. The odd-intrinsic parity sector is related to the chiral anomaly, since the lowest order contribution to this sector comes precisely from the Wess-Zumino-Witten term. For pions alone, its form is fixed by cohomology theory and can be formulated on a 5-dimensional manifold \cite{Witten}. However, the terms that involve external sources can be cast as a four-dimensional integral of chiral invariant densities, {\it{i.e.}}, they are proportional to the Levi-Civit\'a tensor $\varepsilon_{\mu\nu\sigma\rho}$.

In the presence of external sources, the anomalous chiral Lagrangian is known to contribute already at ${\cal{O}}(p^4)$. In \cite{Bijnens:2001bb,Ebertshauser:2001nj} the basis of operators at the next order was determined for vector, axial, scalar and pseudoscalar sources. In the following we will argue that the odd-parity sector involving tensor sources only starts at the $p^8$-order.

In order to obtain the lowest order odd-intrinsic operators in the chiral expansion, the tensor source must have some indices contracted with the
Levi-Civit\'a symbol. In what follows we will show that all such possible contractions identically reduce to even-parity operators. 

Consider first the case when both tensor indices are contracted with the Levi-Civit\'a symbol, {\it{e.g.}},
\begin{equation}
\varepsilon_{\mu\nu\sigma\rho}\,t^{\mu\nu}_{\pm}\,B^{\sigma\rho}\,,
\end{equation}
where $B^{\sigma\rho}$ stands for any tensor structure compatible with chiral and discrete symmetries.
From the definition of the chiral projectors, Eq.~(\ref{projectors}), one can write
\begin{equation}
\varepsilon^{\mu\nu\alpha\beta}\,=\,2\,i\left(P_L^{\mu\nu\alpha\beta}-P_R^{\mu\nu\alpha\beta}\right)\,,
\end{equation}
whence it follows that
\begin{equation}\label{notwo}
\varepsilon_{\mu\nu\alpha\beta}\,t_{\pm}^{\alpha\beta}  \,=\,  2\, i\, t_{\mp\mu\nu}\,,
\end{equation}
and therefore such terms are not present in the odd-intrinsic parity sector.
Notice that this is a consequence of the fact that the tensor source has no chiral partner,
or equivalently that $\gamma_{5}\,\sigma_{\alpha\beta}$ is not an independent Dirac structure.

Consider now the case when only one of the indices of the tensor operator is contracted with
the Levi-Civit\'a density, namely\footnote{All other contractions can be rendered equivalent to this one by means of partial integration.
}
\begin{equation}\label{cont}
\varepsilon_{\mu\nu\alpha\beta}\,t_{\pm}^{\mu\gamma}\,B_{\gamma}^{\,\,\,\nu\alpha\beta}\,,
\end{equation}
where $B_{\gamma}^{\,\,\,\nu\alpha\beta}$ stands for any generic chiral tensor (completely antisymmetric
in $\nu$, $\alpha$ and $\beta$) made out of the elements of Eq.~(\ref{listinv}). We will use the Schouten identity in the form~:
\begin{equation}
g^{\rho\gamma}\varepsilon^{\mu\nu\alpha\beta}-g^{\rho\mu}\varepsilon^{\gamma\nu\alpha\beta}-
g^{\rho\nu}\varepsilon^{\mu\gamma\alpha\beta}-g^{\rho\alpha}\varepsilon^{\mu\nu\gamma\beta}-
g^{\rho\beta}\varepsilon^{\mu\nu\alpha\gamma}=0\,,\label{schouten}
\end{equation} 
which stems from the fact that any 5-form vanishes in 4 dimensions. Contracting it with $t_{\pm}^{\mu\gamma}B_{\gamma}^{\,\,\,\nu\alpha\beta}$ it is not difficult to show (with the use of Eq.~(\ref{notwo})) that it can be rewritten in the following
way~:
\begin{equation}
\varepsilon_{\mu\nu\alpha\beta}\,t_{\pm}^{\mu\gamma}\,B_{\gamma}^{\,\,\,\nu\alpha\beta}\,=\,3\,i\,t_{\mp\alpha\beta}B_{\nu}^{\,\,\,\nu\alpha\beta}\,,\label{eq:loves}\end{equation}
which shows that Eq.~(\ref{cont}) does not contribute to the odd-parity sector.

However, when
none of the indices of the tensor source is contracted with the Levi-Civit\'a density, odd-parity operators will in general arise. If we take, for instance, any of the operators of our
basis at ${\cal{O}}(p^4)$ and multiply it by any of the $\mathcal{O}(p^{4})$ odd-intrinsic
parity operators of Ref.~\cite{Bijnens:2001bb} we will
get an odd-intrinsic parity operator involving tensor currents. But
this operator will be at least of ${\cal{O}}(p^8)$, as anticipated, and it falls beyond the scope of the present work.

\section{On the power counting for the tensor source}
In this section we will elaborate a bit more on our choice for the chiral counting of tensor sources.

Let us begin by briefly reviewing the chiral counting for the other Dirac external fields. In Section 2 we motivated the introduction of external fields coupled to QCD currents as a way to automatically ensure the chiral Ward identities when computing Green's functions. For this to happen, the global chiral symmetry of the QCD Lagrangian has to be promoted to a local one. From the point of view of external fields, this step only affects the vector and axial-vector sources, which play the role of chiral gauge fields and therefore enter the chiral covariant derivative when it replaces the ordinary one. One is then naturally led to make the chiral dimension of the vector and axial-vector sources coincide with that of the ordinary derivative, {\it{i.e.}}, $v_{\mu}, a_{\mu}\sim\mathcal{O}(p)$. Notice that no reference to the actual physical meaning of the sources was needed~: gauge invariance is enough and the sources can be regarded as formal entities.

However, for scalar and pseudoscalar sources the situation changes. In order to motivate their chiral scaling contact has to be made with QCD through quark masses. Quark masses can be formally introduced as external scalar sources, and chiral invariance groups the scalar and pseudoscalar densities in the combination $\chi=2\,B_0(s+i\,p)$ (and its hermitian conjugate), where $B_0$ can be seen as a coupling required by na\"ive dimensional analysis. The Gell-Mann--Oakes--Renner relation for the pion mass sets $m_{\pi}^2=B_0\,(m_u+m_d)$ and one is naturally led to consider $\chi\sim m_{\pi}^2\sim {\cal{O}}(p^2)$. This scaling assignment is of course subject to assuming $B_0\gg F_0$, which seems to be the picture supported by phenomenology. Incidentally, since scalar external sources have a physical realization as the quark masses, the combination $B_0 m_q$ is renormalization invariant and the coupling $B_0$ can be determined by matching $\chi$PT onto the QCD Lagrangian, yielding the well-known result $B_0=-\left\langle {\bar{\psi}}\psi \right\rangle F_0^{-2}$.
 
Therefore, gauge symmetry alone motivates the scaling for vector and axial-vector sources, whereas the momentum scaling for scalars and pseudoscalars is suggested by the way chiral symmetry is (explicitly) broken.

Let us examine the situation for tensor sources. The tensor field coupled to ${\bar{\psi}}\,\sigma_{\mu\nu}\psi$ induces a chirality flip (much like scalars and pseudoscalars do) and therefore transforms in the same way under a chiral transformation. However, unlike scalars and pseudoscalars, tensor fields do not have a physical realization as symmetry breaking terms in the chiral Lagrangian. Their chiral power counting is therefore not motivated by physical arguments and should be seen only as a formal theoretical tool to compute Green's functions. Whatever choice is made for the chiral counting, it will only affect the way operators with different number of tensor sources are organized in the chiral expansion.\footnote{Obviously, the chiral expansion of the different Green's functions with tensor sources is insensitive to the eventual choice of chiral counting for the external fields.} A convenient choice is to assign the tensor source with the same chiral counting as the scalar and pseudoscalar sources, {\it{i.e.}}, ${\bar{t}}_{\mu\nu}\,\sim\,{\cal{O}}(p^2)$. This has two main advantages~: (a) the tensor source only generates even terms in the chiral expansion, and therefore does not change the standard chiral counting scheme; (b) operators involving resonance exchange appear at ${\cal{O}}(p^4)$, leaving only universal terms at ${\cal{O}}(p^2)$.

Since we are assigning the same chiral counting to all spin-flipping sources $s$, $p$ and ${\bar{t}}_{\mu\nu}$, one could equally well define, by analogy to $\chi\,=\,2\,B_0\,(s+ip)$, a tensor chiral field $\tau_{\mu\nu}\,=\,b_0\, {\bar{t}}_{\mu\nu}$. Here $b_0$ would be the analog of $B_0$ for tensor fields. One advantage of introducing a dimensionful parameter $b_0$ is that all the low-energy couplings at a given order in the chiral expansion would then have the same mass dimension. For instance, at ${\mathcal{O}}(p^4)$, the complete set of chiral low energy couplings
\begin{equation}
L_i,\,\,\, i =1,...,10;\qquad H_1, H_2;\qquad \lambda_j,\,\,\, j =1,...,4,
\end{equation}
are dimensionless, where $\lambda_j$ are defined in terms of the $\Lambda_j$ of Eq.~(\ref{O4}) as $\Lambda_j\,=\,b_0^n\,\lambda_j$, $n$ being the number of tensor sources in the associated operators. For instance, $\Lambda_1\,=\,b_0\,\lambda_1$ but $\Lambda_3\,=\,b_0^2\,\lambda_3$.

Furthermore, notice that all operators with external fields we added to the QCD Lagrangian in Eq.~(\ref{sources}) are, by construction, scale invariant. Conservation of vector and axial-vector currents (in the chiral limit) implies in turn that $v_{\mu}$ and $a_{\mu}$ are also scale invariant. On the other hand, the anomalous dimensions of the QCD scalar and pseudoscalar currents are known to be the opposite to that of the quark masses, implying that $s$ and $p$ run like the quark masses. Since the combination $B_0\, m_q$ is scale invariant, it follows that $\chi$ is also invariant. Thus, by analogy, $b_0$ is purported to make $\tau_{\mu\nu}$ renormalization-group invariant.

This can be understood by means of a renormalization group analysis. For the tensor current, 
\begin{equation}
\mu\,\dfrac{\mathrm{d}}{\mathrm{d}\mu}\, T_{\alpha\beta}\,=\,-\,\gamma_{T}\, T_{\alpha\beta}\,,
\end{equation}
where $T_{\alpha\beta}={\bar{\psi}}\,\sigma_{\alpha\beta}\,\psi$ and $\gamma_{T}$ is the tensor anomalous dimension. In the high momentum transfer regime ($\mu\gg\Lambda_{QCD}$), the anomalous dimension can be computed to give
\begin{equation}
\gamma_{T}\,=\,C_{F}\,\dfrac{\alpha_{s}}{2\pi}\,+\,\mathcal{O}(\alpha_{s}^{2})\,.
\end{equation}
Invariance of the QCD Lagrangian implies that the tensor external source ${\bar{t}}_{\mu\nu}$ has to evolve as
\begin{equation}
\mu\,\dfrac{\mathrm{d}}{\mathrm{d}\mu}\, {\bar{t}}_{\alpha\beta}\,=\,\gamma_{T}\, {\bar{t}}_{\alpha\beta}\,.
\end{equation}
Consider now a term in the $\chi$PT Lagrangian with $n$ tensor sources, $\Lambda^{(n)}\, {\cal{O}}_n({\bar{t}}_{\mu\nu})$. When related to QCD parameters, the low energy coupling $\Lambda^{(n)}$ will pick the QCD scale dependence. Defining $\tau_{\mu\nu}\,=\,b_0\, {\bar{t}}_{\mu\nu}$, the term can now be written as $\Lambda^{(n)}\, {\cal{O}}_n({\bar{t}}_{\mu\nu})=\lambda^{(n)}\, {\cal{O}}_n(\tau_{\mu\nu})= (b_0^n\, \lambda^{(n)})\, {\cal{O}}_n({\bar{t}}_{\mu\nu})$. Therefore $\Lambda^{(n)}=b_0^n\,\lambda^{(n)}$ and all the QCD scale dependence is contained in $b_0$, namely
\begin{equation}
\,\mu\,\frac{\mathrm{d}}{\mathrm{d}\mu}\,b_0\,=-\,\gamma_T\,b_0\, .
\end{equation}
This is in complete analogy to the role played by $B_0$ in the scalar-pseudoscalar sector. This analogy can be best illustrated with the following example.
\subsection{A simple application~: One loop corrections to $\mathbf{\Pi_{VT}}$}
Consider the following two-point correlator in the chiral limit
\begin{eqnarray}
\Pi^{VT}_{\mu;\nu\rho}(q)&=&i\int\mathrm{d}^{4}x\, e^{iq\cdot x}\left\langle 0\left|\,T\left\{ V_{\mu}(x)T_{\nu\rho}^{\dagger}(0)\right\} \right|0\right\rangle\nonumber\\
&=&i\,(q^{\rho}g^{\mu\nu}-q^{\nu}g^{\mu\rho})\,\Pi_{VT}(q^2)\, ,
\end{eqnarray}
where $T_{\mu\nu}(x)=\bar{u}(x)\sigma_{\mu\nu}d(x)$ and 
$V_{\mu}(x)=\bar{u}(x)\gamma_{\mu}d(x)$. 
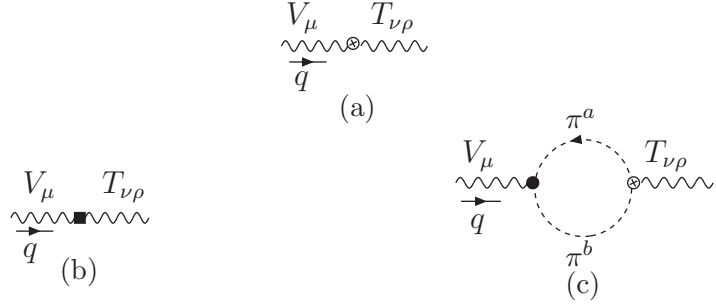
\begin{figure}[t]
\centering

  \begin{picture}(65,35) (325,-250)
    \SetWidth{0.5}
    \SetColor{Black}
    \COval(352,-234)(2.24,2.24)(-26.565052){Black}{White}\Line(352.71,-235.41)(351.29,-232.59)\Line(350.59,-234.71)(353.41,-233.29)
    \ArrowLine(328,-241)(342,-241)
    \Photon(355,-235)(380,-235){2}{4}
    \Photon(325,-235)(350,-235){2}{4}
    \Text(331,-253)[lb]{\large{\Black{$q$}}}
    \Text(328,-231)[lb]{\large{\Black{$V_{\mu}$}}}
    \Text(360,-231)[lb]{\large{\Black{$T_{\nu\rho}$}}}
    \Text(348,-265)[lb]{\normalsize{\Black{(a)}}}
  \end{picture}  
\\
\hskip 1.5cm
\begin{picture}(65,36) (134,-253)
    \SetWidth{0.5}
    \SetColor{Black}
    \Text(169,-233)[lb]{\large{\Black{$T_{\nu\rho}$}}}
    \Text(139,-233)[lb]{\large{\Black{$V_{\mu}$}}}
    \SetWidth{0.5}
    \ArrowLine(136,-243)(150,-243)
    \Text(139,-256)[lb]{\large{\Black{$q$}}}
    \Text(153,-265)[lb]{\normalsize{\Black{(b)}}}
    \CBox(162,-240)(158,-236){Black}{Black}
    \Photon(134,-238)(158,-238){2}{4}
    \Photon(162,-238)(186,-238){2}{4}
  \end{picture}
\hskip 3.5cm
  \begin{picture}(99,64) (135,-190)
    \SetWidth{0.5}
    \SetColor{Black}
    \Photon(135,-162)(162,-162){2}{4}
    \COval(164,-162)(2.24,2.24)(-26.565052){Black}{White}\Line(164.71,-163.41)(163.29,-160.59)\Line(162.59,-162.71)(165.41,-161.29)
    \DashArrowArc(183,-164)(18.25,279,639){2}
    \ArrowLine(137,-169)(152,-169)
    \Vertex(164,-162){2.24}
    \Text(138,-158)[lb]{\large{\Black{$V_{\mu}$}}}
    \Text(207,-158)[lb]{\large{\Black{$T_{\nu\rho}$}}}
    \Text(141,-182)[lb]{\large{\Black{$q$}}}
    \Text(177,-194)[lb]{\normalsize{\Black{$\pi^b$}}}
    \Text(177,-142)[lb]{\normalsize{\Black{$\pi^a$}}}
    \Text(177,-207)[lb]{\normalsize{\Black{(c)}}}
    \COval(202,-162)(2.24,2.24)(-26.565052){Black}{White}\Line(202.71,-163.41)(201.29,-160.59)\Line(200.59,-162.71)(203.41,-161.29)
    \Photon(205,-162)(232,-162){2}{4}
  \end{picture}
\newline
\caption[VT]{Diagrams contributing to (a) the tree level and (b,c) the one-loop renormalization of $\Pi_{VT}(q^2)$. Dotted, square and circle cross vertices correspond, respectively, to ${\cal{O}}(p^2)$, ${\cal{O}}(p^4)$ and ${\cal{O}}(p^6)$ operators in $\chi$PT.}\label{fig1}
\end{figure}

Using dimensional regularization with minimal subtraction, a straightforward computation of the diagrams of Fig.~\ref{fig1} leads to
\begin{equation}
\Pi_{VT}(q^2)\,=\,-\,2\Lambda_1-\Omega_{94}\,q^2+\frac{\Lambda_2}{32\pi^2F_{\pi}^2}\left[\frac{2}{\hat{\epsilon}}-\log{(-q^2)}+\frac{8}{3}\right]q^2\, ,
\end{equation}
where 
\begin{equation}
\frac{2}{\hat{\epsilon}}\,=\,\frac{2}{\epsilon}-\gamma_{E}+\log{4\pi}, \qquad \epsilon=4-d\, .
\end{equation}
In $\chi$PT, renormalization proceeds order by order in the chiral expansion. This means that the logarithmic divergence of Fig.~\ref{fig1}(c) has to be absorbed by the counterterm $\Omega_{94}$ of Fig.~\ref{fig1}(b) to render $\Pi_{VT}(q^2)$ finite. This defines the renormalized coupling $\Omega_{94}^R$ to be
\begin{equation}
\Omega_{94}\,=\,\Omega_{94}^R(\mu)+\frac{\Lambda_2}{16\pi^2F_{\pi}^2}\frac{\mu^{-\epsilon}}{{\hat{\epsilon}}}\, ,\label{94-split}
\end{equation}   
where the chiral scale $\mu$ in $\Omega_{94}^R$ shows the arbitrariness in subtracting the divergence from the bare low energy coupling $\Omega_{94}$. 
The fully renormalized Green's function therefore reads
\begin{equation}
\Pi_{VT}(q^2)\,=\,-\,2\,\Lambda_1-\Omega_{94}^R(\mu)\,q^2+\frac{\Lambda_2}{32\pi^2F_{\pi}^2}\left[\frac{8}{3}-\log{\left(-\frac{q^2}{\mu^2}\right)}\right]q^2\, .
\end{equation}
So far, the scale dependence associated to the tensor current has been implicitly stored into $\Lambda_1$, $\Omega_{94}$ and $\Lambda_2$. If we now introduce the aforementioned parameter $b_0$, we find $\Lambda_{1,2}=b_0\,\lambda_{1,2}$ and $\Omega_{94}=b_0\,\omega_{94}$ and as a result 
\begin{equation}\label{VTb0}
\Pi_{VT}(q^2)\,=\,-\,2\,\lambda_1\,b_0-\omega_{94}^R(\mu)\,b_0\,q^2+\frac{\lambda_2\,b_0}{32\pi^2F_{\pi}^2}\left[\frac{8}{3}-\log{\left(-\frac{q^2}{\mu^2}\right)}\right]q^2\, .
\end{equation} 
Notice that with the $b_0$ parameter, the chiral scale and the QCD scale factorize.

For comparison consider now the following scalar-pseudoscalar two-point Green's function~:
\begin{equation}
\Pi_{SS-PP}(q)\,=\,i\int\mathrm{d}^{4}x\, e^{iq\cdot x}\left\langle 0\left|\,T\left\{ S(x)S^{\dagger}(0)-P(x)P^{\dagger}(0)\right\} \right|0\right\rangle\, ,
\end{equation}
where $S(x)={\bar{u}}(x)d(x)$ and $P(x)={\bar{u}}(x)i\gamma_5d(x)$. After evaluating the corresponding Feynman diagrams, one obtains, in the chiral limit,
\begin{equation}
\Pi_{SS-PP}(q^2)\,=\,\frac{2F_0^2\,B_0^2}{q^2}+32B_0^2\,L_8+\frac{5\,B_0^2}{48\pi^2}\left[\frac{2}{{\hat{\epsilon}}}-\log{(-q^2)}+2\right]\, .
\end{equation}
Again, the previous equation determines the (chiral) scale dependence of the renormalized coupling~:
\begin{equation}
L_8\,=\,L_8^R(\mu)-\frac{5}{48}\frac{1}{16\pi^2}\frac{\mu^{-\epsilon}}{{\hat{\epsilon}}}\, ,\label{L8-split}
\end{equation}
leading to the one-loop renormalized two-point Green's function
\begin{equation}
\Pi_{SS-PP}(q^2)\,=\,\frac{2F_0^2\,B_0^2}{q^2}+32B_0^2\,L_8^R(\mu)+\frac{5\,B_0^2}{48\pi^2}\left[ 2-\log{\left(\frac{-q^2}{\mu^2}\right)}\right]\, .
\end{equation} 
All the QCD scale dependence, arising from the non-conservation of the scalar and pseudoscalar currents, is factored out in $B_0$, whereas $L_8^R(\mu)$ shows the running with the chiral scale. Notice the analogy with Eq.~(\ref{VTb0}).

Unfortunately, $b_0$ cannot be matched onto the QCD Lagrangian in a way similar to what is done for $B_0$~: the lowest dimension operators linear in the tensor source (and consequently in $b_0$) are coupled to the low-energy couplings $\Lambda_1$ and $\Lambda_2$. These couplings are insensitive to pion dynamics and instead do receive contributions from vector-meson resonances \cite{CMinprogress}. Therefore, there is an inherent ambiguity in the determination of $b_0$, because it cannot be decoupled from $\Lambda_1$ and $\Lambda_2$. The dimensionful coupling $b_0$ should not contain information on the integrated degrees of freedom of the theory, but otherwise it remains unspecified. To avoid confusion, we have omitted in our treatment any reference to $b_0$.

As a result, one should keep in mind that, besides the chirally renormalized low-energy couplings, each operator with $n$ tensor sources bears a non-vanishing anomalous dimension, namely $n\,\gamma_T$.

\section{Conclusions}
We have built the most general $C,P$ and chiral invariant Lagrangian to $\mathcal{O}(p^4)$ and $\mathcal{O}(p^6)$ including the sources coupled to the tensor Dirac bilinear current ${\bar{\psi}}\sigma_{\mu\nu}\psi$. We have assigned the tensor sources with a chiral counting such as to preserve the original scheme of even terms in the chiral expansion. In order to end up with a minimal set of operators use has been made of the leading-order equations of motion, integration by parts and the Bianchi identity. Specialization to $n_f=2,3$ provides additional relations by the use of the Cayley-Hamilton theorem, the full set of which are listed in the Appendix. For the three-flavor case one finds 110 new operators and 3 contact terms, while for two flavors one ends up with 75 operators and 3 contact terms. We have also shown that operators contributing to the odd-intrinsic parity sector with tensor fields start not before ${\cal{O}}(p^8)$.

\section*{Acknowledgements}

We would like to thank S.~Peris, A.~Pich, J.~Portol\'es and I.~Rosell for reading the manuscript and providing useful suggestions. O.~C. is also grateful to the University of Washington for hospitality and to S.~R.~Sharpe for fruitful discussions at the different stages of this work. We thank C.~Haefeli, M.~A.~Ivanov and M.~Schmid for drawing our attention to the results of \cite{Haefeli:2007ty} and W.~Detmold for pointing out the work done with tensor operators in generalised parton distributions (GPDs). V.~M. thanks C.~Haefeli for useful discussions about minimality of the basis. The work of O.~C. is supported by the Fulbright Program and the Spanish Ministry of Education and Science under grant no. FU2005-0791. The work of V.~M. is supported by a FPU contract (MEC). This work has been supported in part by the EU MRTN-CT-2006-035482 (FLAVIAnet), by MEC (Spain) under grant FPA2004-00996 and by Generalitat Valenciana under grant GVACOMP2007-156.

\section*{Appendix~: Cayley-Hamilton relations}
\newcounter{alpha}
\renewcommand{\thesection}{\Alph{alpha}}
\renewcommand{\theequation}{\Alph{alpha}.\arabic{equation}}
\renewcommand{\thetable}{\Alph{alpha}}
\setcounter{alpha}{1}
\setcounter{equation}{0}
\setcounter{table}{0}
The analysis in the main text to build the basis of operators has dealt with  general $SU(n_f)$. In practice, however, one wants to specialize to the phenomenologically relevant cases, $n_f=2$ and $n_f=3$. The Cayley-Hamilton theorem states that any square $n\times n$ matrix $A$ satisfies its own characteristic 
equation, $\chi_n(A)=0$. This sets a relation between $A$ and their invariants (traces and determinant). The form of the relation depends on the dimensionality $n$ of the linear space. For instance,
\begin{eqnarray}
\chi_2(A)=A^2-\langle A\rangle A+({\mathrm{det}}A)\,{\bf{1}}_{2\times 2}&=&0,\qquad (n=2);\nonumber\\
\chi_3(A)=A^3-\langle A\rangle A^2+\frac{\langle A\rangle^2-\langle A^2\rangle}{2}A-({\mathrm{det}}A)\,{\bf{1}}_{3\times 3}&=&0,\qquad (n=3).
\end{eqnarray}
One immediate consequence of the previous equations is that the determinant of any matrix is a function of its traces. We have implicitly used this information to write all chiral invariants solely in terms of traces. Solving the previous equations for the determinant, one finds
\begin{eqnarray}
A^2-\langle A\rangle A+\frac{\langle A\rangle^2-\langle A^2\rangle}{2}\,{\bf{1}}_{2\times 2}&=&0\,,\nonumber\\
A^3-\langle A\rangle A^2+\frac{\langle A\rangle^2-\langle A^2\rangle}{2}A-\left[\frac{\langle A^3\rangle}{3}-\frac{\langle A^2\rangle\langle A\rangle}{2}-\frac{\langle A\rangle^3}{6}\right]\,{\bf{1}}_{3\times 3}&=&0\,.
\end{eqnarray}
Cayley-Hamilton relations therefore set constraints between traces. For these constraints to be non-trivial, one has to build relations involving at least $(n+1)$ matrices. For instance, for $n=2$ the quantity $\langle a\,\chi_2(b+c)\rangle$ gives
\begin{equation}\label{C-H}
\langle a\{b,c\}\rangle -\langle a\rangle \langle bc\rangle -\langle b\rangle \langle ca\rangle 
-\langle c\rangle \langle ab\rangle +\langle a\rangle \langle b\rangle\langle c\rangle=0\,, 
\end{equation}
whereas for $n=3$, using $\langle a\,\chi_3(b+c+d)\rangle$, one ends up with
\begin{eqnarray}\label{C-H-3}
&\langle ab\{c,d\}\rangle+
\langle ac\{b,d\}\rangle+
\langle ad\{b,c\}\rangle-
\langle a\{b,c\}\rangle\langle d\rangle-
\langle a\{b,d\}\rangle\langle c\rangle-
\langle a\{c,d\}\rangle\langle b\rangle-\nonumber\\
&-\langle b\{c,d\}\rangle\langle a\rangle-
\langle ab\rangle \langle cd\rangle-
\langle ac\rangle \langle bd\rangle-
\langle ad\rangle \langle bc\rangle+
\langle a\rangle \langle b\rangle\langle cd\rangle+
\langle a\rangle \langle c\rangle\langle bd\rangle+\nonumber\\
&+\langle a\rangle \langle d\rangle\langle bc\rangle+
\langle b\rangle \langle c\rangle\langle ad\rangle+
\langle b\rangle \langle d\rangle\langle ac\rangle+
\langle c\rangle \langle d\rangle\langle ab\rangle-
\langle a\rangle \langle b\rangle\langle c\rangle\langle d\rangle=0\,. 
\end{eqnarray}
After imposing the Cayley-Hamilton relations, in Table \ref{list2} we have favored the terms with a minimum number of traces, bearing in mind that these are the dominant ones in a large-$N_c$ expansion of the chiral Lagrangian.

\subsection{$\mathbf{SU(3)}$}
For $n_f=3$, use of Eq.~(\ref{C-H-3}) leads to the following relations,
\begin{eqnarray*}
i\left\langle t_{+\mu\nu}\left[u^{\mu},u^{\alpha}\right]\right\rangle \left\langle u_{\alpha}u^{\nu}\right\rangle [Y_{6}] & = & Y_{1}+2\,Y_{3}+Y_{4}-Y_{5},\\
i\left\langle t_{+\mu\nu}u_{\alpha}\right\rangle \left\langle u^{\alpha}u^{\mu}u^{\nu}\right\rangle [Y_{7}] & = & Y_{1}+Y_{2}-\frac{1}{2}\,Y_{5}-Y_{8},\\
\left\langle t_{+\mu\nu}\right\rangle \left\langle t_{+}^{\mu\nu}\right\rangle \left\langle u_{\alpha}u^{\alpha}\right\rangle [Y_{21}] & = &- 4\,Y_{9}-2\,Y_{10}+Y_{14}+2\,Y_{16}+4\,Y_{19},\\
\left\langle t_{+\mu\nu}\right\rangle \left\langle t_{+}^{\mu\alpha}\right\rangle \left\langle u_{\alpha}u^{\nu}\right\rangle [Y_{22}] & = & -2\,Y_{11}-2\,Y_{12}-Y_{13}+Y_{15}+Y_{17}+Y_{18}+2\,Y_{20},\\
\left\langle t_{-\mu\nu}\right\rangle \left\langle t_{-}^{\mu\alpha}\right\rangle \left\langle u_{\alpha}u^{\nu}\right\rangle [Y_{30}] & = & -2\,Y_{23}-2\,Y_{24}-Y_{25}+Y_{26}+Y_{27}+Y_{28}+2\,Y_{29},\\
\left\langle t_{+\mu\nu}u_{\alpha}\right\rangle \left\langle f_{+}^{\mu\nu}u^{\alpha}\right\rangle [Y_{64}] & = & Y_{57}+Y_{58}-\frac{1}{2}\,Y_{62}-Y_{67},\\
\left\langle t_{+\mu\nu}u_{\alpha}\right\rangle \left\langle f_{+}^{\mu\alpha}u^{\nu}\right\rangle [Y_{65}] & = & Y_{59}+Y_{60}+Y_{61}-Y_{63}-Y_{66}-Y_{68}.\end{eqnarray*}
\subsection{$\mathbf{SU(2)}$}

The relations derived in the previous section also hold for two flavors. In addition, repetitive use of Eq.~(\ref{C-H}) can be used to reduce monomials with multiple traces containing at least three chiral operators. We find

\allowdisplaybreaks
\begin{eqnarray*}
i\left\langle t_{+\mu\nu}\left\{ u_{\alpha}u^{\alpha},u^{\mu}u^{\nu}\right\} \right\rangle [Y_{1}] & = & 2\,Y_{3},\\
i\left\langle t_{+\mu\nu}\left\{ u_{\alpha},u^{\mu}u^{\alpha}u^{\nu}\right\} \right\rangle [Y_{4}] & = & -Y_{2}-Y_{3},\\
i\left\langle t_{+\mu\nu}u^{\mu}u^{\nu}\right\rangle \left\langle u_{\alpha}u^{\alpha}\right\rangle [Y_{5}] & = & 2\,Y_{3},\\
i\left\langle t_{+\mu\nu}\right\rangle \left\langle u_{\alpha}u^{\alpha}u^{\mu}u^{\nu}\right\rangle [Y_{8}] & = & 0,\\
\left\langle t_{+\mu\nu}t_{+}^{\mu\nu}\right\rangle \left\langle u_{\alpha}u^{\alpha}\right\rangle [Y_{14}] & = & 2\,Y_{9},\\
\left\langle t_{+\mu\nu}t_{+}^{\mu\alpha}\right\rangle \left\langle u^{\nu}u_{\alpha}\right\rangle [Y_{15}] & = & Y_{11}+Y_{12},\\
\left\langle t_{+\mu\nu}u^{\alpha}\right\rangle \left\langle t_{+}^{\mu\alpha}u_{\alpha}\right\rangle [Y_{16}] & = & Y_{9}+Y_{10}-Y_{19},\\
\left\langle t_{+\mu\nu}u^{\alpha}\right\rangle \left\langle t_{+}^{\mu\alpha}u_{\nu}\right\rangle [Y_{17}] & = & Y_{12}+\frac{1}{2}\,Y_{13}-\frac{1}{2}\,Y_{20},\\
\left\langle t_{+\mu\nu}u^{\nu}\right\rangle \left\langle t_{+}^{\mu\alpha}u_{\alpha}\right\rangle [Y_{18}] & = & Y_{11}+\frac{1}{2}\,Y_{13}-\frac{1}{2}\,Y_{20},\\
\left\langle t_{-\mu\nu}t_{-}^{\mu\alpha}\right\rangle \left\langle u^{\nu}u_{\alpha}\right\rangle [Y_{26}] & = & Y_{23}+Y_{24},\\
\left\langle t_{-\mu\nu}u^{\alpha}\right\rangle \left\langle t_{-}^{\mu\alpha}u_{\nu}\right\rangle [Y_{27}] & = & Y_{24}+\frac{1}{2}\,Y_{25}-\frac{1}{2}\,Y_{29},\\
\left\langle t_{-\mu\nu}u^{\nu}\right\rangle \left\langle t_{-}^{\mu\alpha}u_{\alpha}\right\rangle [Y_{28}] & = & Y_{23}+\frac{1}{2}\,Y_{25}-\frac{1}{2}\,Y_{29},\\
\left\langle t_{+\mu\nu}\right\rangle \left\langle t_{+}^{\mu\nu}\right\rangle \left\langle \chi_{+}\right\rangle [Y_{37}] & = & -2\,Y_{31}+Y_{33}+2\,Y_{34},\\
\left\langle t_{-\mu\nu}\right\rangle \left\langle t_{+}^{\mu\nu}\right\rangle \left\langle \chi_{-}\right\rangle [Y_{38}] & = & -2\,Y_{32}+Y_{35}+2\,Y_{36},\\
i\left\langle \chi_{+}\right\rangle \left\langle t_{+\mu\nu}u^{\mu}u^{\nu}\right\rangle [Y_{41}] & = & \frac{1}{2}\,Y_{39}+Y_{40},\\
i\left\langle t_{+\mu\nu}\right\rangle \left\langle \chi_{+}u^{\mu}u^{\nu}\right\rangle [Y_{42}] & = & \frac{1}{2}\,Y_{39}-Y_{40},\\
i\left\langle \chi_{-}\right\rangle \left\langle t_{-\mu\nu}u^{\mu}u^{\nu}\right\rangle [Y_{45}] & = & \frac{1}{2}\,Y_{43}+Y_{44},\\
i\left\langle t_{-\mu\nu}\right\rangle \left\langle \chi_{-}u^{\mu}u^{\nu}\right\rangle [Y_{46}] & = & \frac{1}{2}\,Y_{43}-Y_{44},\\
\left\langle t_{+\mu\nu}f_{+}^{\mu\nu}\right\rangle \left\langle u_{\alpha}u^{\alpha}\right\rangle [Y_{62}] & = & Y_{57},\\
\left\langle t_{+\mu\nu}f_{+}^{\mu\alpha}\right\rangle \left\langle u^{\nu}u_{\alpha}\right\rangle [Y_{63}] & = & \frac{1}{2}\left(Y_{59}+Y_{60}\right),\\
\left\langle t_{+\mu\nu}u^{\nu}\right\rangle \left\langle f_{+}^{\mu\alpha}u_{\alpha}\right\rangle [Y_{66}] & = & \frac{1}{2}\left(Y_{60}+Y_{61}\right),\\
\left\langle t_{+\mu\nu}\right\rangle \left\langle f_{+}^{\mu\nu}u_{\alpha}u^{\alpha}\right\rangle [Y_{67}] & = & 0,\\
\left\langle t_{+\mu\nu}\right\rangle \left\langle f_{+}^{\mu\alpha}\left\{u_{\alpha},u^{\nu}\right\}\right\rangle [Y_{68}] & = & 0,\\
\left\langle t_{+\mu\nu}f_{+}^{\mu\nu}\right\rangle \left\langle \chi_{+}\right\rangle [Y_{79}] & = & Y_{74}-Y_{77},\\
\left\langle t_{-\mu\nu}f_{+}^{\mu\nu}\right\rangle \left\langle \chi_{-}\right\rangle [Y_{80}] & = & Y_{75}-Y_{78},\\
i\left\langle t_{-}^{\nu\rho}\right\rangle \left\langle t_{+\mu\nu}h_{\,\,\rho}^{\mu}\right\rangle [Y_{83}] & = & Y_{81}-Y_{82},\\
i\left\langle t_{-\mu\nu}\right\rangle \left\langle f_{-}^{\nu\rho}f_{+\rho}^{\mu}\right\rangle [Y_{87}] & = & Y_{86},\\
i\left\langle t_{-}^{\nu\rho}\right\rangle \left\langle f_{-\mu\nu}t_{+\rho}^{\mu}\right\rangle [Y_{93}] & = & Y_{90}-Y_{91}-Y_{93}-4\,Y_{119},\\
i\left\langle \partial^{\mu}t_{-\mu\nu}\right\rangle \left\langle f_{+}^{\nu\rho}u_{\rho}\right\rangle [Y_{101}] & = & Y_{98},\\
i\left\langle \partial_{\rho}t_{-\mu\nu}\right\rangle \left\langle f_{+}^{\mu\nu}u^{\rho}\right\rangle [Y_{102}] & = & Y_{99},\\
i\left\langle \partial_{\rho}t_{-\mu\nu}\right\rangle \left\langle f_{+}^{\mu\rho}u^{\nu}\right\rangle [Y_{103}] & = & Y_{100},\\
i\left\langle t_{+\mu\nu}\right\rangle \left\langle \nabla_{\lambda}t_{-}^{\mu\nu}u^{\lambda}\right\rangle [Y_{109}] & = & Y_{104}-Y_{106},\\
i\left\langle t_{+\nu\lambda}\right\rangle \left\langle \nabla_{\mu}t_{-}^{\mu\nu}u^{\lambda}\right\rangle [Y_{110}] & = & \frac{1}{2}\,Y_{11}+\frac{1}{4}\,Y_{13}-\frac{1}{2}\,Y_{23}-\frac{1}{4}\,Y_{25}+Y_{52}-Y_{53}+Y_{105}-Y_{107}-4\,Y_{118},\\
i\left\langle t_{-\nu\lambda}\right\rangle \left\langle \nabla_{\mu}t_{+}^{\mu\nu}u^{\lambda}\right\rangle [Y_{111}] & = & Y_{105}-Y_{108},\\
i\left\langle t_{-\mu\nu}\right\rangle \left\langle h^{\nu\rho}f_{+\rho}^{\mu}\right\rangle [Y_{117}] & = & Y_{116}.\end{eqnarray*}

%

\newpage
\renewcommand{\arraystretch}{1.3}\setlength{\LTcapwidth}{\textwidth}

\begin{longtable}[c]{|l|c|c|c|c|}

\hline
\multicolumn{1}{|c|}{monomial $Y_{i}$}&
$SU(n_f)$&
$SU(3)$&
$SU(2)$\\
\hline
\hline
\endhead
\hline
\caption[List of operators contributing to the $\mathcal{O}(p^6)$ Lagrangian.]{\rule{0cm}{2em}List of operators contributing to the $\mathcal{O}(p^6)$ Lagrangian.}
\endfoot
\hline
\caption[List of operators contributing to the $\mathcal{O}(p^6)$ Lagrangian.]{\label{list2}
\rule{0cm}{2em}List of operators contributing to the $\mathcal{O}(p^6)$ Lagrangian.}
\endlastfoot
$i\left\langle t_{+\mu\nu}\left\{ u_{\alpha}u^{\alpha},u^{\mu}u^{\nu}\right\} \right\rangle $&
1&
1&
\\
$i\left\langle t_{+\mu\nu}u^{\alpha}u^{\mu}u^{\nu}u_{\alpha}\right\rangle $&
2&
2&
1\\
$i\left\langle t_{+\mu\nu}u^{\mu}u_{\alpha}u^{\alpha}u^{\nu}\right\rangle $&
3&
3&
2\\
$i\left\langle t_{+\mu\nu}\left\{ u_{\alpha},u^{\mu}u^{\alpha}u^{\nu}\right\} \right\rangle $&
4&
4&
\\
$i\left\langle t_{+\mu\nu}u^{\mu}u^{\nu}\right\rangle \left\langle u_{\alpha}u^{\alpha}\right\rangle $&
5&
5&
\\
$i\left\langle t_{+\mu\nu}\left[u^{\mu},u^{\alpha}\right]\right\rangle \left\langle u_{\alpha}u^{\nu}\right\rangle $&
6&
&
\\
$i\left\langle t_{+\mu\nu}u_{\alpha}\right\rangle \left\langle u^{\alpha}u^{\mu}u^{\nu}\right\rangle $&
7&
&
\\
$i\left\langle t_{+\mu\nu}\right\rangle \left\langle u_{\alpha}u^{\alpha}u^{\mu}u^{\nu}\right\rangle $&
8&
6&
\\
&
&
&
\\
$\left\langle t_{+\mu\nu}t_{+}^{\mu\nu}u_{\alpha}u^{\alpha}\right\rangle $&
9&
7&
3\\
$\left\langle t_{+\mu\nu}u_{\alpha}t_{+}^{\mu\nu}u^{\alpha}\right\rangle $&
10&
8&
4\\
$\left\langle t_{+\mu\nu}t_{+}^{\mu\alpha}u_{\alpha}u^{\nu}\right\rangle $&
11&
9&
5\\
$\left\langle t_{+\mu\nu}t_{+}^{\mu\alpha}u^{\nu}u_{\alpha}\right\rangle $&
12&
10&
6\\
$\left\langle t_{+\mu\nu}\left(u^{\nu}t_{+}^{\mu\alpha}u_{\alpha}+u_{\alpha}t_{+}^{\mu\alpha}u^{\nu}\right)\right\rangle $&
13&
11&
7\\
$\left\langle t_{+\mu\nu}t_{+}^{\mu\nu}\right\rangle \left\langle u_{\alpha}u^{\alpha}\right\rangle $&
14&
12&
\\
$\left\langle t_{+\mu\nu}t_{+}^{\mu\alpha}\right\rangle \left\langle u^{\nu}u_{\alpha}\right\rangle $&
15&
13&
\\
$\left\langle t_{+\mu\nu}u_{\alpha}\right\rangle \left\langle t_{+}^{\mu\nu}u^{\alpha}\right\rangle $&
16&
14&
\\
$\left\langle t_{+\mu\nu}u_{\alpha}\right\rangle \left\langle t_{+}^{\mu\alpha}u^{\nu}\right\rangle $&
17&
15&
\\
$\left\langle t_{+\mu\nu}u^{\nu}\right\rangle \left\langle t_{+}^{\mu\alpha}u_{\alpha}\right\rangle $&
18&
16&
\\
$\left\langle t_{+\mu\nu}\right\rangle \left\langle t_{+}^{\mu\nu}u_{\alpha}u^{\alpha}\right\rangle $&
19&
17&
8\\
$\left\langle t_{+\mu\nu}\right\rangle \left\langle t_{+}^{\mu\alpha}\left\{ u_{\alpha},u^{\nu}\right\} \right\rangle $&
20&
18&
9\\
$\left\langle t_{+\mu\nu}\right\rangle \left\langle t_{+}^{\mu\nu}\right\rangle \left\langle u_{\alpha}u^{\alpha}\right\rangle $&
21&
&
\\
$\left\langle t_{+\mu\nu}\right\rangle \left\langle t_{+}^{\mu\alpha}\right\rangle \left\langle u_{\alpha}u^{\nu}\right\rangle $&
22&
&
\\
&
&
&
\\
$\left\langle t_{-\mu\nu}t_{-}^{\mu\alpha}u_{\alpha}u^{\nu}\right\rangle $&
23&
19&
10\\
$\left\langle t_{-\mu\nu}t_{-}^{\mu\alpha}u^{\nu}u_{\alpha}\right\rangle $&
24&
20&
11\\
$\left\langle t_{-\mu\nu}\left(u^{\nu}t_{-}^{\mu\alpha}u_{\alpha}+u_{\alpha}t_{-}^{\mu\alpha}u^{\nu}\right)\right\rangle $&
25&
21&
12
\\
$\left\langle t_{-\mu\nu}t_{-}^{\mu\alpha}\right\rangle \left\langle u^{\nu}u_{\alpha}\right\rangle $&
26&
22&
\\
$\left\langle t_{-\mu\nu}u_{\alpha}\right\rangle \left\langle t_{-}^{\mu\alpha}u^{\nu}\right\rangle $&
27&
23&
\\
$\left\langle t_{-\mu\nu}u^{\nu}\right\rangle \left\langle t_{-}^{\mu\alpha}u_{\alpha}\right\rangle $&
28&
24&
\\
$\left\langle t_{-\mu\nu}\right\rangle \left\langle t_{-}^{\mu\alpha}\left\{ u_{\alpha},u^{\nu}\right\} \right\rangle $&
29&
25&
13\\
$\left\langle t_{-\mu\nu}\right\rangle \left\langle t_{-}^{\mu\alpha}\right\rangle \left\langle u_{\alpha}u^{\nu}\right\rangle $&
30&
&
\\
&
&
&
\\
$\left\langle t_{+\mu\nu}t_{+}^{\mu\nu}\chi_{+}\right\rangle $ &
31&
26&
14\\
$\left\langle t_{+\mu\nu}t_{-}^{\mu\nu}\chi_{-}\right\rangle $&
32&
27&
15\\
$\left\langle t_{+\mu\nu}t_{+}^{\mu\nu}\right\rangle \left\langle \chi_{+}\right\rangle $&
33&
28&
16\\
$\left\langle t_{+\mu\nu}\chi_{+}\right\rangle \left\langle t_{+}^{\mu\nu}\right\rangle $&
34&
29&
17\\

%
$\left\langle t_{+\mu\nu}t_{-}^{\mu\nu}\right\rangle \left\langle \chi_{-}\right\rangle $&
35&
30&
18\\
$\left\langle t_{+\mu\nu}\chi_{-}\right\rangle \left\langle t_{-}^{\mu\nu}\right\rangle $&
36&
31&
19\\
$\left\langle t_{+\mu\nu}\right\rangle \left\langle t_{+}^{\mu\nu}\right\rangle \left\langle \chi_{+}\right\rangle $&
37&
32&
\\
$\left\langle t_{+\mu\nu}\right\rangle \left\langle t_{-}^{\mu\nu}\right\rangle \left\langle \chi_{-}\right\rangle $&
38&
33&
\\
&
&
&
\\
$i\left\langle t_{+\mu\nu}\left\{ \chi_{+},u^{\mu}u^{\nu}\right\} \right\rangle $&
39&
34&
20\\
$i\left\langle t_{+\mu\nu}u^{\mu}\chi_{+}u^{\nu}\right\rangle $&
40&
35&
21\\
$i\left\langle \chi_{+}\right\rangle \left\langle t_{+\mu\nu}u^{\mu}u^{\nu}\right\rangle $&
41&
36&
\\
$i\left\langle t_{+\mu\nu}\right\rangle \left\langle \chi_{+}u^{\mu}u^{\nu}\right\rangle $&
42&
37&
\\
$i\left\langle t_{-\mu\nu}\left\{ \chi_{-},u^{\mu}u^{\nu}\right\} \right\rangle $&
43&
38&
22\\
$i\left\langle t_{-\mu\nu}u^{\mu}\chi_{-}u^{\nu}\right\rangle $&
44&
39&
23\\
$i\left\langle \chi_{-}\right\rangle \left\langle t_{-\mu\nu}u^{\mu}u^{\nu}\right\rangle $&
45&
40&
\\
$i\left\langle t_{-\mu\nu}\right\rangle \left\langle \chi_{-}u^{\mu}u^{\nu}\right\rangle $&
46&
41&
\\
&
&
&
\\
$\left\langle t_{-\mu\nu}\left(h^{\nu\rho}u_{\rho}u^{\mu}-u^{\mu}u_{\rho}h^{\nu\rho}\right)\right\rangle $&
47&
42&
24\\
$\left\langle t_{-\mu\nu}\left(h^{\nu\rho}u^{\mu}u_{\rho}-u_{\rho}u^{\mu}h^{\nu\rho}\right)\right\rangle $&
48&
43&
25\\
$\left\langle t_{-\mu\nu}\left(u_{\rho}h^{\nu\rho}u^{\mu}-u^{\mu}h^{\nu\rho}u_{\rho}\right)\right\rangle $&
49&
44&
26\\
$\left\langle t_{-\mu\nu}\right\rangle \left\langle h^{\nu\rho}\left[u_{\rho},u^{\mu}\right]\right\rangle $&
50&
45&
27\\
&
&
&
\\
$\left\langle \nabla_{\rho}t_{+\mu\nu}\nabla^{\rho}t_{+}^{\mu\nu}\right\rangle $&
51&
46&
28\\
$\left\langle \nabla_{\mu}t_{+}^{\mu\nu}\nabla^{\rho}t_{+\rho\nu}\right\rangle $&
52&
47&
29\\
$\left\langle \nabla_{\mu}t_{-}^{\mu\nu}\nabla^{\rho}t_{-\rho\nu}\right\rangle $&
53&
48&
30\\
$\left\langle \partial_{\rho}t_{+\mu\nu}\right\rangle \left\langle \partial^{\rho}t_{+}^{\mu\nu}\right\rangle $&
54&
49&
31\\
$\left\langle \partial^{\mu}t_{+\nu\mu}\right\rangle \left\langle \partial^{\rho}t_{+\rho}^{\nu}\right\rangle $&
55&
50&
32\\
$\left\langle \partial^{\mu}t_{-\nu\mu}\right\rangle \left\langle \partial^{\rho}t_{-\rho}^{\nu}\right\rangle $&
56&
51&
33\\
&
&
&
\\
$\left\langle t_{+\mu\nu}\left\{ f_{+}^{\mu\nu},u_{\alpha}u^{\alpha}\right\} \right\rangle $&
57&
52&
34\\
$\left\langle t_{+\mu\nu}u_{\alpha}f_{+}^{\mu\nu}u^{\alpha}\right\rangle $&
58&
53&
35\\
$\left\langle t_{+\mu\nu}\left(f_{+}^{\mu\alpha}u^{\nu}u_{\alpha}+u_{\alpha}u^{\nu}f_{+}^{\mu\alpha}\right)\right\rangle $&
59&
54&
36\\
$\left\langle t_{+\mu\nu}\left(f_{+}^{\mu\alpha}u_{\alpha}u^{\nu}+u^{\nu}u_{\alpha}f_{+}^{\mu\alpha}\right)\right\rangle $&
60&
55&
37\\
$\left\langle t_{+\mu\nu}\left(u^{\nu}f_{+}^{\mu\alpha}u_{\alpha}+u_{\alpha}f_{+}^{\mu\alpha}u^{\nu}\right)\right\rangle $&
61&
56&
38\\
$\left\langle t_{+\mu\nu}f_{+}^{\mu\nu}\right\rangle \left\langle u_{\alpha}u^{\alpha}\right\rangle $&
62&
57&
\\
$\left\langle t_{+\mu\nu}f_{+}^{\mu\alpha}\right\rangle \left\langle u^{\nu}u_{\alpha}\right\rangle $&
63&
58&
\\
$\left\langle t_{+\mu\nu}u_{\alpha}\right\rangle \left\langle f_{+}^{\mu\nu}u^{\alpha}\right\rangle $&
64&
&
\\
$\left\langle t_{+\mu\nu}u_{\alpha}\right\rangle \left\langle f_{+}^{\mu\alpha}u^{\nu}\right\rangle $&
65&
&
\\
$\left\langle t_{+\mu\nu}u^{\nu}\right\rangle \left\langle f_{+}^{\mu\alpha}u_{\alpha}\right\rangle $&
66&
59&
\\
$\left\langle t_{+\mu\nu}\right\rangle \left\langle f_{+}^{\mu\nu}u_{\alpha}u^{\alpha}\right\rangle $&
67&
60&
\\
$\left\langle t_{+\mu\nu}\right\rangle \left\langle f_{+}^{\mu\alpha}\left\{ u_{\alpha},u^{\nu}\right\} \right\rangle $&
68&
61&
\\
&
&
&
\\
$\left\langle t_{-\mu\nu}\left[f_{-}^{\mu\nu},u_{\alpha}u^{\alpha}\right]\right\rangle $&
69&
62&
39\\
$\left\langle t_{-\mu\nu}\left(f_{-}^{\mu\alpha}u^{\nu}u_{\alpha}-u_{\alpha}u^{\nu}f_{-}^{\mu\alpha}\right)\right\rangle $&
70&
63&
40\\

%
$\left\langle t_{-\mu\nu}\left(f_{-}^{\mu\alpha}u_{\alpha}u^{\nu}-u^{\nu}u_{\alpha}f_{-}^{\mu\alpha}\right)\right\rangle $&
71&
64&
41\\
$\left\langle t_{-\mu\nu}\left(u^{\nu}f_{-}^{\mu\alpha}u_{\alpha}-u_{\alpha}f_{-}^{\mu\alpha}u^{\nu}\right)\right\rangle $&
72&
65&
42\\
$\left\langle t_{-\mu\nu}\right\rangle \left\langle f_{-}^{\mu\alpha}\left[u_{\alpha},u^{\nu}\right]\right\rangle $&
73&
66&
43\\
$\left\langle t_{+\mu\nu}\left\{ f_{+}^{\mu\nu},\chi_{+}\right\} \right\rangle $ &
74&
67&
44\\
$\left\langle t_{-\mu\nu}\left\{ f_{+}^{\mu\nu},\chi_{-}\right\} \right\rangle $ &
75&
68&
45\\
$\left\langle t_{+\mu\nu}\left[f_{-}^{\mu\nu},\chi_{-}\right]\right\rangle $ &
76&
69&
46\\
$\left\langle t_{+\mu\nu}\right\rangle \left\langle f_{+}^{\mu\nu}\chi_{+}\right\rangle $&
77&
70&
47\\
$\left\langle t_{-\mu\nu}\right\rangle \left\langle f_{+}^{\mu\nu}\chi_{-}\right\rangle $&
78&
71&
48\\
$\left\langle t_{+\mu\nu}f_{+}^{\mu\nu}\right\rangle \left\langle \chi_{+}\right\rangle $&
79&
72&
\\
$\left\langle t_{-\mu\nu}f_{+}^{\mu\nu}\right\rangle \left\langle \chi_{-}\right\rangle $&
80&
73&
\\
&
&
&
\\
$i\left\langle t_{+\mu\nu}\left\{ t_{-}^{\nu\rho},h_{\,\,\rho}^{\mu}\right\} \right\rangle $&
81&
74&
49\\
$i\left\langle t_{+\mu\nu}\right\rangle \left\langle t_{-}^{\nu\rho}h_{\,\,\rho}^{\mu}\right\rangle $&
82&
75&
50\\
$i\left\langle t_{-}^{\nu\rho}\right\rangle \left\langle t_{+\mu\nu}h_{\,\,\rho}^{\mu}\right\rangle $&
83&
76&
\\
&
&
&
\\
$i\left\langle t_{+\mu\nu}f_{-}^{\mu\rho}f_{-\rho}^{\nu}\right\rangle $&
84&
77&
51\\
$i\left\langle t_{+\mu\nu}f_{+}^{\mu\rho}f_{+\rho}^{\nu}\right\rangle $&
85&
78&
52\\
$i\left\langle t_{-\mu\nu}\left\{ f_{-}^{\nu\rho},f_{+\rho}^{\mu}\right\} \right\rangle $&
86&
79&
53\\
$i\left\langle t_{-\mu\nu}\right\rangle \left\langle f_{-}^{\nu\rho}f_{+\rho}^{\mu}\right\rangle $&
87&
80&
\\
&
&
&
\\
$i\left\langle t_{+\mu\nu}t_{+}^{\mu\rho}t_{+\rho}^{\nu}\right\rangle $&
88&
81&
54\\
$i\left\langle t_{+\mu\nu}t_{-}^{\mu\rho}t_{-\rho}^{\nu}\right\rangle $&
89&
82&
55\\
&
&
&
\\
$i\left\langle f_{+\mu\nu}t_{+}^{\nu\rho}t_{+\rho}^{\mu}\right\rangle $&
90&
83&
56\\
$i\left\langle f_{+\mu\nu}t_{-}^{\nu\rho}t_{-\rho}^{\mu}\right\rangle $&
91&
84&
57\\
$i\left\langle t_{+\rho}^{\mu}\right\rangle \left\langle f_{-\mu\nu}t_{-}^{\nu\rho}\right\rangle $&
92&
85&
58\\
$i\left\langle t_{-}^{\nu\rho}\right\rangle \left\langle f_{-\mu\nu}t_{+\rho}^{\mu}\right\rangle $&
93&
86&
\\
&
&
&
\\
$\left\langle \nabla_{\mu}t_{+}^{\mu\nu}\nabla^{\alpha}f_{+\alpha\nu}\right\rangle $&
94&
87&
59\\
&
&
&
\\
$i\left\langle \nabla_{\rho}t_{+\mu\nu}\left[h^{\mu\rho},u^{\nu}\right]\right\rangle $&
95&
88&
60\\
$i\left\langle \nabla^{\mu}t_{+\mu\nu}\left[h^{\nu\rho},u_{\rho}\right]\right\rangle $&
96&
89&
61\\
$i\left\langle \nabla^{\mu}t_{+\mu\nu}\left[f_{-}^{\nu\rho},u_{\rho}\right]\right\rangle $&
97&
90&
62\\
$i\left\langle \nabla^{\mu}t_{-\mu\nu}\left\{ f_{+}^{\nu\rho},u_{\rho}\right\} \right\rangle $&
98&
91&
63\\
$i\left\langle \nabla_{\rho}t_{-\mu\nu}\left\{ f_{+}^{\mu\nu},u^{\rho}\right\} \right\rangle $&
99&
92&
64\\
$i\left\langle \nabla_{\rho}t_{-\mu\nu}\left\{ f_{+}^{\mu\rho},u^{\nu}\right\} \right\rangle $&
100&
93&
65\\
$i\left\langle \partial^{\mu}t_{-\mu\nu}\right\rangle \left\langle f_{+}^{\nu\rho}u_{\rho}\right\rangle $&
101&
94&
\\
$i\left\langle \partial_{\rho}t_{-\mu\nu}\right\rangle \left\langle f_{+}^{\mu\nu}u^{\rho}\right\rangle $&
102&
95&
\\
$i\left\langle \partial_{\rho}t_{-\mu\nu}\right\rangle \left\langle f_{+}^{\mu\rho}u^{\nu}\right\rangle $&
103&
96&
\\

%
$i\left\langle \left\{ \nabla_{\lambda}t_{-}^{\mu\nu},t_{+\mu\nu}\right\} u^{\lambda}\right\rangle $&
104&
97&
66\\
$i\left\langle \left\{ \nabla_{\mu}t_{+}^{\mu\nu},t_{-\nu\lambda}\right\} u^{\lambda}\right\rangle $&
105&
98&
67\\
$i\left\langle \partial_{\lambda}t_{-}^{\mu\nu}\right\rangle \left\langle t_{+\mu\nu}u^{\lambda}\right\rangle $&
106&
99&
68\\
$i\left\langle \partial_{\mu}t_{-}^{\mu\nu}\right\rangle \left\langle t_{+\nu\lambda}u^{\lambda}\right\rangle $&
107&
100&
69\\
$i\left\langle \partial_{\mu}t_{+}^{\mu\nu}\right\rangle \left\langle t_{-\nu\lambda}u^{\lambda}\right\rangle $&
108&
101&
70\\
$i\left\langle t_{+\mu\nu}\right\rangle \left\langle \nabla_{\lambda}t_{-}^{\mu\nu}u^{\lambda}\right\rangle $&
109&
102&
\\
$i\left\langle t_{+\nu\lambda}\right\rangle \left\langle \nabla_{\mu}t_{-}^{\mu\nu}u^{\lambda}\right\rangle $&
110&
103&
\\
$i\left\langle t_{-\nu\lambda}\right\rangle \left\langle \nabla_{\mu}t_{+}^{\mu\nu}u^{\lambda}\right\rangle $&
111&
104&
\\
&
&
&
\\
$\left\langle t_{-}^{\mu\nu}\left[\chi_{+\mu},u_{\nu}\right]\right\rangle $&
112&
105&
71\\
$\left\langle t_{+}^{\mu\nu}\left[\chi_{-\mu},u_{\nu}\right]\right\rangle $&
113&
106&
72\\
&
&
&
\\
$i\left\langle t_{+\mu\nu}h^{\mu\alpha}h_{\,\,\alpha}^{\nu}\right\rangle $&
114&
107&
73\\
&
&
&
\\
$i\left\langle t_{+\mu\nu}\left[h^{\mu\alpha},f_{-\alpha}^{\nu}\right]\right\rangle $&
115&
108&
74\\
$i\left\langle t_{-\mu\nu}\left\{ h^{\mu\alpha},f_{+\alpha}^{\nu}\right\} \right\rangle $&
116&
109&
75\\
$i\left\langle t_{-\mu\nu}\right\rangle \left\langle h^{\mu\alpha}f_{+\alpha}^{\nu}\right\rangle $&
117&
110&
\\
&
&
&
\\
&
&
&
\\
\hline
Contact terms&
&
&\\
&
&
&
\\
$\left\langle D_{\mu}t^{\mu\nu}D^{\alpha}t_{\alpha\nu}^{\dagger}\right\rangle $&
118&
111&
76\\
$i\left\langle t^{\dagger\nu\rho}t^{\mu}_{\,\,\rho}F_{L\mu\nu}+t^{\nu\rho}t_{\,\,\,\,\,\rho}^{\dagger\mu}F_{R\mu\nu}\right\rangle $&
119&
112&
77\\
$\left\langle t_{\mu\nu}\chi^{\dagger}F_{R}^{\mu\nu}+\chi t_{\mu\nu}^{\dagger}F_{R}^{\mu\nu}+t_{\mu\nu}^{\dagger}\chi F_{L}^{\mu\nu}+\chi^{\dagger}t_{\mu\nu}F_{L}^{\mu\nu}\right\rangle $&
120&
113&
78\\
\end{longtable}

\end{document}